\documentclass[letterpaper,twocolumn,10pt]{article}

\usepackage{usenix,epsfig,endnotes}
\usepackage{verbatim}
\usepackage{booktabs}
\usepackage{tabularx}
\usepackage{subcaption}
\usepackage{enumitem}
\usepackage{xspace}
\usepackage{multirow}
\usepackage{soul, color, xcolor}
\usepackage{multirow}
\usepackage{listings}
\usepackage{amsmath}
\usepackage{amsfonts}
\usepackage{amssymb}
\usepackage{booktabs}
\usepackage{url}
\usepackage[linesnumbered,ruled,vlined]{algorithm2e} 
\usepackage{stfloats}
\usepackage{etoolbox}
\usepackage{marvosym}
\usepackage{appendix}

\newcommand{\squishlist}{
 \begin{list}{$\bullet$}
  { \setlength{\itemsep}{0pt}
     \setlength{\parsep}{3pt}
     \setlength{\topsep}{3pt}
     \setlength{\partopsep}{0pt}
     \setlength{\leftmargin}{1.5em}
     \setlength{\labelwidth}{1em}
     \setlength{\labelsep}{0.5em} } }

\newcommand{\squishlisttwo}{
 \begin{list}{$\bullet$}
  { \setlength{\itemsep}{0pt}
     \setlength{\parsep}{0pt}
    \setlength{\topsep}{0pt}
    \setlength{\partopsep}{0pt}
    \setlength{\leftmargin}{2em}
    \setlength{\labelwidth}{1.5em}
    \setlength{\labelsep}{0.5em} } }

\newcommand{\squishend}{
  \end{list}  }

\newbool{showComments}
\booltrue{showComments}

\ifbool{showComments}{%
\newcommand{\zhenyu}[1]{\sethlcolor{orange}\hl{[ZY: #1]}}
\newcommand{\fy}[1]{\sethlcolor{pink}\hl{[Furong: #1]}}
\newcommand{\jxadd}[1]{\sethlcolor{white}\hl{#1}}
\newcommand{\jxdel}[1]{\sethlcolor{pink}\hl{}}
}{

\newcommand{\zhenyu}[1]{} 
\newcommand{\fy}[1]{}
\newcommand{\jxadd}[1]{}
\newcommand{\jxdel}[1]{}
}

\begin{document}
\date{}


\title{ \Large \bf Choir: Tackling RTBC Performance Impossible Triangle with 5G Collaboration}

\author{
{\rm Wenji Du}$^{1,2}$,
{\rm Wanghong Yang}$^{2}$,
{\rm Baosen Zhao}$^{1,2}$,
{\rm Yongmao Ren}$^{2}$,
{\rm Xu Zhou}$^{2}$,\\
{\rm Jiaxing Zhang}$^{1,3}$,
{\rm Tingting Yuan}$^{4}$,
{\rm Qinghua Wu}$^{3}$,
{\rm Xiaoming Fu}$^{4}$,
{\rm Gaogang Xie}$^{1,2}$\\[0.5em]
$^{1}$University of Chinese Academy of Sciences, China \quad
$^{2}$Computer Network Information Center, CAS, China\\
$^{3}$Institute of Computing Technology, CAS, China \quad
$^{4}$University of G\"ottingen, Germany
}

\maketitle

\subsection*{Abstract}
Real-time Broadband Communication (RTBC) scenarios, such as cloud virtual reality and 8K live streaming, further raise the criteria of the \textit{Performance Triangle}, including video bitrates exceeding 30Mbps, tail delay below 50ms, and fairness guarantees for multi-user concurrent access.
Based on our testing and analysis, existing RTBC-oriented rate control solutions, including end-to-end algorithms and network-assisted algorithms, fail to simultaneously satisfy all performance metrics.
The native dynamic delay and physical layer resource allocation strategy inherent to 5G radio access network (RAN) are the key reasons. 
These solutions lack adaptation to the 5G architecture, leading to reduced decision performance.

This paper proposes Choir, an innovative collaborative solution mainly deployed on 5G base stations that deeply integrates 5G radio characteristics and video streaming traffic patterns to guide sender-side efficient rate control.
Extensive simulation and testbed evaluations demonstrate Choir's significant performance on high average bitrate, low tail delay and inter-flow fairness over different 5G network scenarios.

\section{Introduction}
Compared to real-time communication (RTC) scenarios primarily pursuing low latency, real-time broadband communication (RTBC) scenarios require the simultaneous satisfaction of the \textit{Performance Triangle} defined as ultra-low latency, ultra-high throughput, and fairness.
Taking cloud virtual reality in densely populated entertainment spaces as an example, delivering an immersive user experience requires latency below 50 ms and a bitrate of at least 30 Mbps. 
The fairness issue also becomes increasingly prominent. 
Users urgently demand stable transmission unaffected by concurrent users within the same access network\cite{torres2020immersive,vrwhitepaper,rubio2017immersive,kamarainen2017measurement,7938641,singh2013performance}.

\begin{figure}[tbp]
    \centering
    \includegraphics[width=1\linewidth]{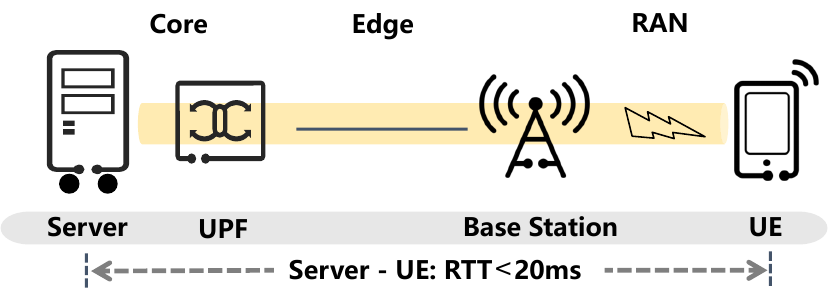}
    \caption{Transmission scenario of RTBC services in 5G}
    \label{fig:scenario}
\end{figure}

With broad bandwidth and robust mobility support, 5G radio access networks (RAN) serve as a critical foundation for RTBC services. 
As shown in Fig. ~\ref{fig:scenario}, RTBC servers can be deployed near the core data processing unit, User Plane Function (UPF), in 5G networks, ensuring that the basic link propagation delays with the user equipment (UE) does not exceed 20ms\cite{3gpp-release-16}. 

Researchers have proposed numerous rate control solutions for low-latency video streaming transmission \cite{johansson2017scream, arun2018copa, ray2022sqp,wang2024pudica, goyal2020abc, l4s, zhuge, shi-scone-rtc-requirement-02}. 
These solutions typically evaluate network states via end-side active probing or network nodes feedback, then design video bitrate and sending rate control algorithms to rapidly respond to network fluctuations, preventing network congestion and latency spikes.

However, these rate control solutions fail to meet RTBC user requirements in 5G networks, even though some algorithms demonstrate good performance in Ethernet and Wi-Fi.

\textbf{The underlying architecture of 5G RAN is the root cause.}

Even without packet loss, the Time Division Duplex (TDD) pattern at the physical layer inherently subjects packets to dynamic transmission delays due to alternating uplink/downlink transmission periods. 

When the radio channel’s Block Error Rate (BLER) increases, retransmissions —from the Hybrid Automatic Repeat reQuest (HARQ) mechanism at the Media Access Control (MAC) layer to the Acknowledged Mode (AM) at the Radio Link Control (RLC) layer—introduce additional latency up to tens of milliseconds.

These RAN-induced dynamic delays force existing rate control solutions into persistent oscillation, failing to fully utilize RAN capacity while struggling to achieve stable low latency, which is unacceptable for RTBC users.

Conversely, 5G RAN’s per-flow dedicated queues and priority-aware scheduling inherently guarantee physical-layer resource fairness. 
By enabling rate control solutions to rapidly adapt to dynamic resource allocations, this architecture maybe tackle the \textit{Performance Triangle}: simultaneously maximizing bandwidth utilization, minimizing queuing delay, and ensuring inter-flow fairness.

We propose \textbf{Choir}, a comprehensive bandwidth prediction and rate control solution that jointly accounts for 5G RAN’s dynamic resource allocation mechanisms and video traffic characteristics, guides sender-side bitrate and pacing rate adaptation, thereby achieving optimal bitrate-delay performance across diverse channel conditions while guaranteeing fairness among concurrent flows.

We implemented and evaluated Choir with several state-of-the-art (SOTA) rate control solutions on our 5G module simulation and open-source 5G platform. Experimental results show that Choir reduces tail latency by 68.64\% to 90.7\% in various 5G network scenarios.

In summary, our key contributions in this paper are:
\begin{itemize}
    \item We tested and analyzed the performance of existing SOTA RTC rate control solutions in 5G RAN network, with fine-grained delay breakdown.
    \item We proposed \textbf{Choir}, an innovative collaborative solution that enable base stations to perform accurate guidance bandwidth prediction which dynamically mapped from physical-layer resource allocations with incorporating video traffic characteristics, and synchronizing the guidance bandwidth with end-side rate control algorithms timely.
    \item We deployed and extensively evaluated Choir in trace-driven simulation and open-source 5G testbed, demonstrating its superior performance to achieve the \textit{Performance Triangle}. 
\end{itemize}

\section{Background and Motivation}
\subsection{Tradeoff Strategy of Existing RTBC Solutions}
Existing solutions aim to achieve the \textit{Performance Triangle} but often have to make tradeoff decisions among the three metrics. 

\textbf{Latency}, especially tail latency, is prioritized since long tail latency causes playback stalls that rapidly degrade user retention. 

\textbf{Bitrate} ranks second, as it directly impacts visual quality; for immersive devices like VR, reduced quality triggers user motion sickness.

Regarding \textbf{fairness}, heterogeneous flows (e.g., BBR\cite{bbr} competing with Cubic\cite{cubic}) struggle to achieve stable fairness due to conflicting algorithmic goals. 
Thus, fairness is typically a secondary consideration after ensuring latency and bitrate performance, and it is frequently compromised.

Even if prioritizing low latency and high bitrate without strict fairness requirements, solutions still face two challenges: accurate network state estimation and timely response to network dynamics. 
These challenges have spurred numerous ideas, generally categorized into two types: end-side active probing or network-assisted feedback.

\begin{figure}[tbp]
    \centering
    \includegraphics[width=1\linewidth]{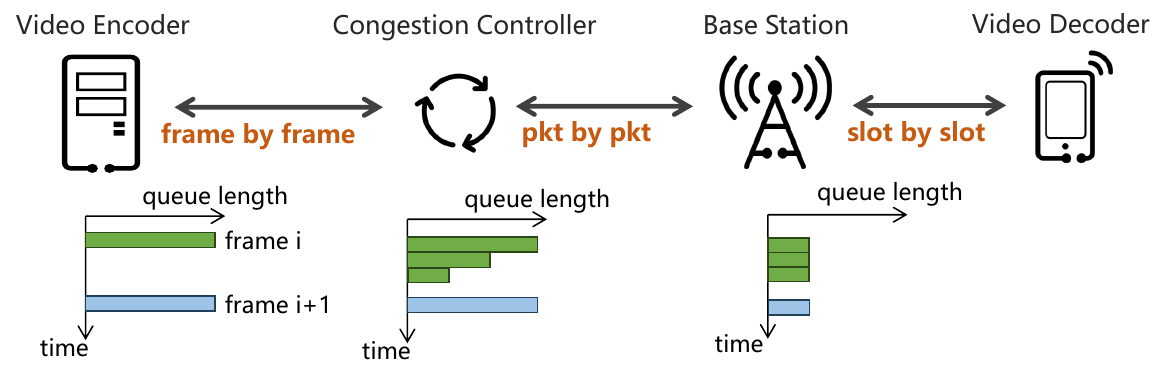}
    \caption{The transmission procedure of video stream in 5G.}
    \label{fig:pacing}
\end{figure}

As shown in Fig.~\ref{fig:pacing}, these solutions fundamentally operate at the sender side, integrating bitrate and framerate control mechanisms for video encoders with congestion control algorithms (CCAs) in pacing controller.

End-side solutions primarily infer network metric changes through data packet and returned ACK transmission states. Examples include: SCReAM \cite{johansson2017scream}, COPA \cite{arun2018copa} and SQP \cite{ray2022sqp} are window-based, which rely on packet trains signals; Pudica \cite{wang2024pudica} sends probing packets to measure bandwidth utilization; PBE-CC \cite{xie2020pbe} extracts and feeds back wireless channel states at the receiver. 

Network-assisted solutions leverage network nodes to explicitly report bottleneck capacity and congestion.

Explicit Congestion Notification (ECN) \cite{ramakrishnan2001addition}, ABC \cite{goyal2020abc}, and L4S \cite{son2024adaptable, l4s, briscoe2019implementing} signal congestion via in-band packet header markings. Their feedback signals must be processed at the receiver, which would be lagged by a longer feedback loop during congestion.
Zhuge\cite{zhuge} implicitly notifies senders via delayed ACKs, its access-point-assisted method (e.g., direct uplink scheduling control) saves congestion signal propagation delays;
Recent RTC proposal SCONE\cite{shi-scone-rtc-requirement-02} proactively feed back bottleneck bandwidth and queueing length for video streams.

\textbf{Since most RTC solutions take delay as a key congestion signal, accurately detecting true network states becomes increasingly challenging in 5G RANs characterized by delay fluctuations.}

\begin{figure}[!ht]
   \centering
   \begin{subfigure}{1\linewidth}
       \centering
       \includegraphics[width=\linewidth]{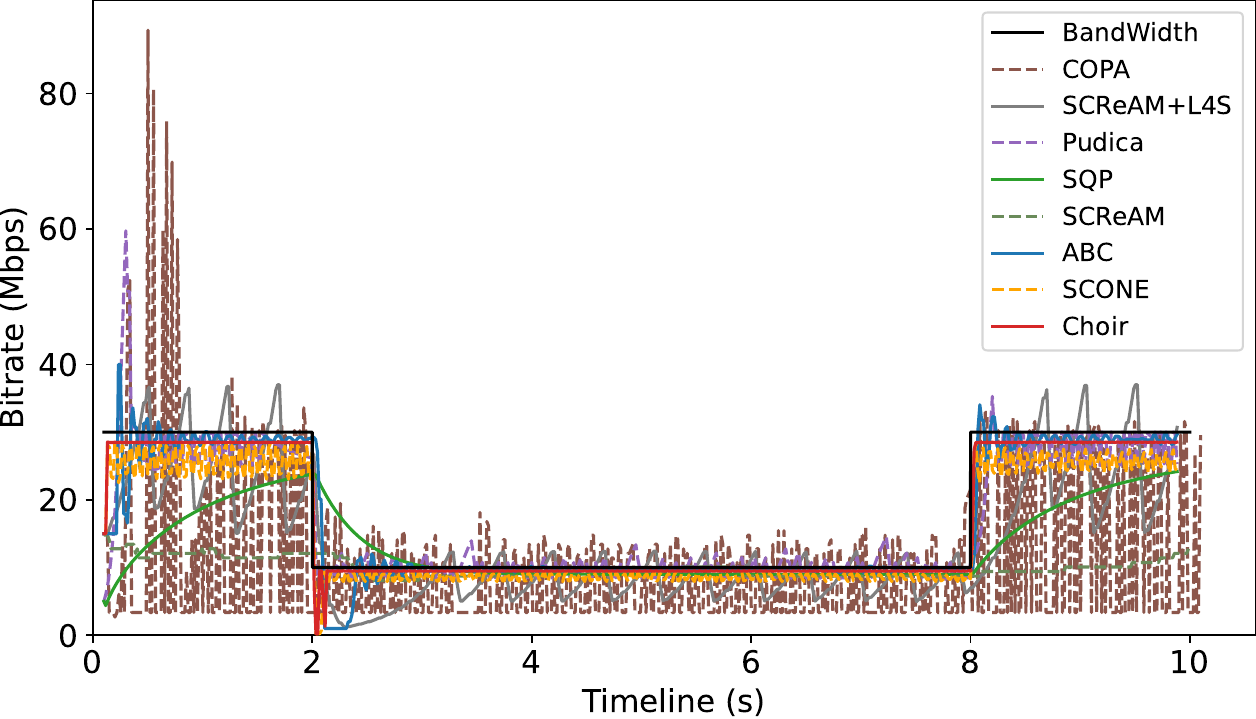} 
       \caption{Bitrate}
       \label{fig:image1}
   \end{subfigure}
   \begin{subfigure}{1\linewidth}
       \centering
       \includegraphics[width=\linewidth]{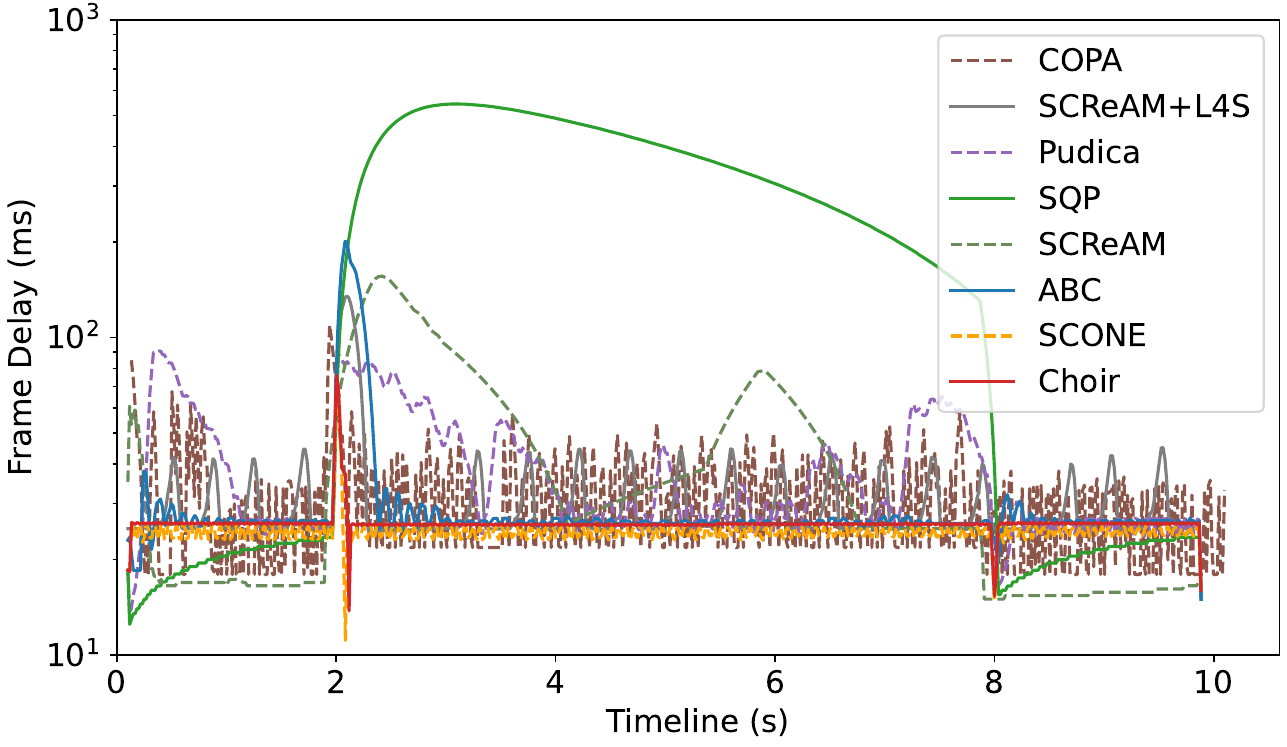} 
       \caption{Frame delay}
       \label{fig:image2}
   \end{subfigure}
   \caption{Performance of different solutions in 5G networks.}
   \label{fig:first_compare}
\end{figure}

Our experiment results, as shown in Fig.~\ref{fig:first_compare}, indicate that most solutions (COPA, Pudica, SQP, SCReAM, SCONE) exhibit unstable transmission rates and persistent delay oscillations in even stable 5G RAN environments; furthermore, abrupt bandwidth reductions exacerbate sustained queue accumulation and elevated tail latency.

Although ABC and L4S exhibit quick adaptation to bandwidth reduction, they require a long convergence time and continue to experience rate fluctuations even after convergence. 

\subsection{Challenges Raised by 5G RAN: Native Dynamic Delay}

The inherent design of 5G RAN at the physical, MAC, and RLC layers introduces native dynamic latency, fundamentally hindering stable convergence for SOTA solutions targeting low latency.

\textbf{Physical Layer Dynamics:}
5G RAN configures radio subframes with TDD patterns, dividing physical channels into alternating downlink (D) and uplink (U) periods. 
For example, the widely adopted DDDSU pattern (designed for heavy downlink traffic) allocates downlink-to-uplink slots in a 3.5:1.5 ratio. 
Data packets arriving in the downlink queue during uplink slots may incur additional queuing delays (e.g., ~1.5 ms). 
Consequently, even in stable networks, packets of identical sizes experience variable transmission delays depending on their arrival timing.

\textbf{Retransmission-Induced Latency:}
To mask packet loss from upper layers and senders, 5G implements retransmission mechanisms at both MAC and RLC layers, given its high channel BLER (~10\% in our tests).
\begin{itemize}
    \item MAC Layer: First HARQ retransmission introduces ~6 ms of latency, accumulating to ~16 ms after three retransmissions.
    \item RLC Layer: In RLC AM, retransmitted packets re-enter the RLC buffer queue, where delays depend on queue length and may exceed tens of milliseconds.
\end{itemize}
While this design aims to prevent sender-side rate reduction triggered by perceived packet loss, it inadvertently amplifies tail latency in practice.

RTBC video streams are disproportionately affected by retransmission mechanisms. 
Even when transmitted over unreliable protocols like UDP, subsequent frames experience cascading queuing delays during RAN channel fluctuations, as underlying retransmissions of prior frames occupy shared buffer resources, leading to multi-frame latency escalation.

Traditional solutions, such as congestion ratio signaling or end-side probing, prove insufficient to promptly and accurately diagnose the root causes of latency spikes in 5G RAN. 
Only solutions like SCONE, which directly reports base station statuss including available bandwidth and queue length, achieve relatively stable performance.

\subsection{Opportunity Provided by 5G RAN: Native Resource Fairness}
\label{opportunity}

However, the 5G RAN architecture also provides physical foundation for the realization of the \textit{Performance Triangle}.

In Wi-Fi and wired networks, routers typically manage multiple same-priority flows within a single queue, leading to inevitable inter-flow competition. 
A single flow’s excessive data volume can degrade transmission latency for all co-existing flows, triggering network-wide congestion.

In contrast, 5G RAN implements per-flow dedicated queues and a polling-based scheduling mechanism for same-priority multi-flow scenarios, achieving \textbf{physical resource fairness}. 
Consequently, each flow’s rate control mechanism neither requires nor can monopolize other users’ physical bandwidth.

This aligns with the Congestion Control Algorithm Independence (CCAI) principle proposed by \cite{brown2024principles}, where CCAs are theoretically relieved from enforcing fairness, thereby focusing solely on achieving bandwidth-delay joint optimization.

However, unlike persistent flows, the intermittent transmission of video frames causes queues of multiple video streams within the same base station to rapidly toggle between active and inactive states, resulting in more frequent resource allocation fluctuations.

To prevent queue buildup and prolonged delay, sender-side reactions must synchronize with physical-layer capacity allocation at ultra-low latency. 
Existing solutions fail to adapt to such rapid resource dynamics. 
Moreover, divergent response logics across solutions further exacerbate throughput imbalance among concurrent flows.

Under these conditions, the target of rate control solutions shift toward minimizing latency while elastically utilizing base station-allocated capacity, thereby satisfying RTBC users’ dual Quality-of-Experience (QoE) demands for high bitrates and low latency.

\textbf{We propose assigning both network dynamics estimation and traffic pattern-aware rate adaptation to the base station.} 

This architectural shift enables RTBC senders within the same RAN to achieve max-min fair resource utilization across concurrent flows through minimal communication overhead, realizing the latency-minimal transmission at higher bitrates. 
To achieve this vision, we design a base station-collaborated rate control framework specifically for 5G RAN’s unique concurrency and channel dynamics, \textbf{Choir}.

\section{Choir Design}
\subsection{Overview}

\begin{figure*}[htbp]
    \centering
    \includegraphics[width=0.9\linewidth]{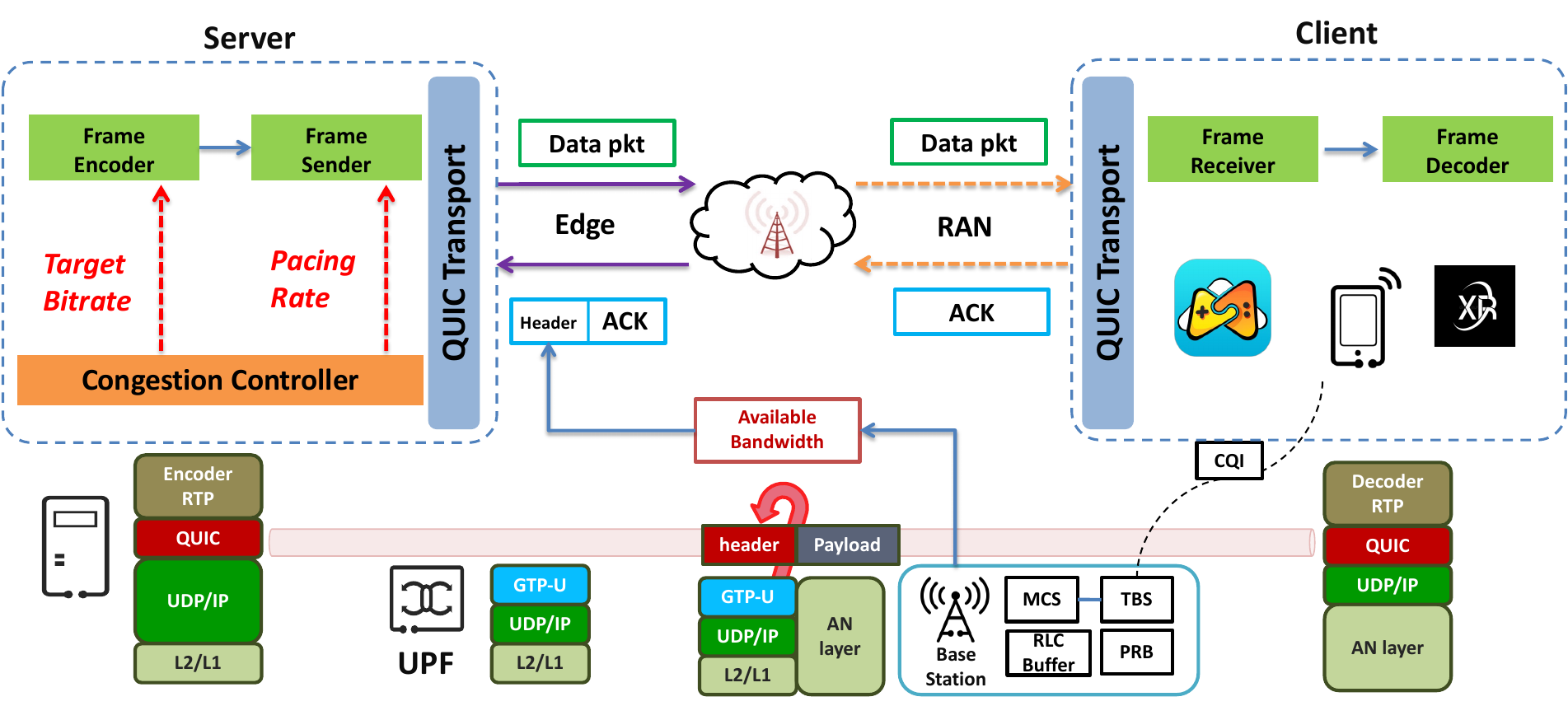}
    \caption{The Framework of Choir}
    \label{fig:framework3}
\end{figure*}

The overall framework of Choir is shown in Fig.~\ref{fig:framework3}. 
To achieve the \textit{Performance Triangle} goals for RTBC video streaming in 5G networks, Choir must precisely capture the dynamics of 5G RAN and video stream characteristics for accurate rate control.

Choir implements two key functions at the base station. 
First, it estimates the allocated bandwidth for each queue in the next scheduling period using underlying channel information. 
Second, it predicts queue length trends by analyzing video stream patterns, calculates draining rates targeting zero queuing, derives available bandwidth, and delivers this guidance value to senders via uplink ACK headers.

Additionally, since existing sender congestion controllers cannot recognize or utilize the guidance bandwidth, Choir deploys a lightweight, QoE-oriented rate control algorithm on the sender side. This algorithm converts guidance bandwidth into target bitrate and pacing rate.

\subsection{Radio characteristics oriented capacity estimation}
This module ensures precise mapping from physical resource allocation to actual allocated bandwidth for each flow's queue at base station.

\textit{\textbf{\underline{Challenge 1:}} To accurately capture highly dynamic resource allocation, bandwidth prediction must operate at the granularity of the minimum scheduling period.}

In mobile networks, multiple flows with the same priority frequently transition between active and inactive states, forcing the base station to perform high-frequency resource reallocation.
Thus, the base station first evaluates resource allocation for each user in the next transmission slot, which is generally 0.5ms or 1ms in 5G RAN.

At the beginning of each connection, this module calculates the number of available physical resource blocks (PRBs) to each flow $i$ based on the current number of concurrent active flows $N_{active}$.

\begin{equation}
    PRB_{i} = \frac{PRB_{RAN}}{N_{active}} 
    \label{eq5}
\end{equation}

\noindent where $PRB_{RAN}$ represents the total number of PRBs of the current base station.
$PRB_{i}$ represents the number of PRBs that flow $i$ can use per transmission time interval (TTI) of the base station. 
This value stabilizes when the number of concurrent active flows remains constant.

For a new TTI, base station need to update $PRB_{i}$ according to the idle resource in the last TTI. 
We then define $PRB_{used}$ as the total number of PRBs allocated by the base station to all active users and $uPRB_{i}$ as the number of used PRBs of active flow $i$ in the last TTI. 
$PRB_{used}$ should not exceed $PRB_{RAN}$. Thus,
\begin{equation}
    PRB_{i} = max(uPRB_{i}+\frac{PRB_{RAN}-PRB_{used}}{N_{active}},\frac{PRB_{RAN}}{N_{total}})
    \label{eq7}
\end{equation}

The base station derives the modulation and coding scheme (MCS) value based on the channel quality indicator (CQI) reported by the UE to determine the bit rate per symbol. 
Meanwhile, the scheduling algorithm considers the current network load to allocate resources, especially the transport block size (TBS) per TTI, to optimize network performance.
Thus, the physical layer transmission rate $PR_{i}$ per PRB for the current flow $i$ in bytes/ms is given by:
 
\begin{equation}
    PR_{i} = \frac{TBS_{i}}{{TTI} \cdot PRB_{i}} \cdot \frac{DN}{TN}
    \label{eq6}
\end{equation}

\noindent where $TBS_{i}$ represents the number of bytes sent by flow $i$ in 1 TTI at the physical layer of the base station.

The uplink period length (UL) and downlink period length (DL) are determined by the radio frame structure of 5G base station. 
We calculate relevant parameters by measuring the total number of TTI ($TN$) and the total number of downlink TTI ($DN$) over a 100ms period.

It is important to note that not all downlink TTI are used for transmitting user data; some may be allocated for control signaling and other purposes. 
Therefore, when counting $DN$, we only consider those actually used for user data transmission, ensuring that the measurements more accurately reflect the downlink capacity.

Therefore, the theoretical maximum physical capacity $C_{i}$ for flow $i$ in the next TTI can be determined as:
\begin{equation}
    C_{i} = PRB_{i} \cdot PR_{i}
    \label{eq8}
\end{equation}

\textit{\textbf{\underline{Challenge 2:}} Mapping physical layer capacity to transport layer bandwidth requires accounting for MAC and RLC layer retransmissions that occupy additional bandwidth.}

We define  $R_{ReTx}$ represents the retransmission rate.
On a shorter timescale, the $R_{ReTx}$ can be calculated as the number of HARQ ($HN$) to the total downlink TTI ($DN$) over a given period.

To improve accuracy, this module selects 10ms as the short-term statistical period.
Compared to the physical layer parameter block error rate (BLER), which is statistically updated only every 100 ms, $R_{ReTx}$ provides higher temporal granularity.
\begin{equation}
    R_{ReTx} =\frac{HN}{DN}
    \label{eq9bw1}
\end{equation}

Thus, this module can calculate the allocated bandwidth for flow $i$ in the next TTI, denoted as $BW_i$:
\begin{equation}
    BW_{i} =\frac{ C_{i} \cdot \gamma  }{1+R_{ReTx}}
    \label{eq9bw2}
\end{equation}

\noindent where $\gamma$ represents the proportion of effective data in a physical layer transport block after removing the overhead. 
Since the data size of each transport block varies, the overheader ratio also differs. 
Therefore, this proportion is calculated by averaging the data across all transport blocks within a 10ms window. 

\subsection{Flow pattern oriented available bandwidth prediction}
This module integrates video stream transmission dynamics to map allocated bandwidth $BW_{i}$ into near-zero queue guidance bandwidth $\widehat{BW_{i}}$ for the sender.

In pursuit of low latency, the general formulation for guidance bandwidth $\widehat{BW_{i}}$ can be expressed as:
\begin{equation}
    \widehat{BW_{i}}= (\eta \times \overline {BW_{i}} - DR_{i})^{+}
    \label{eq10bwnew}
\end{equation}

$\eta$ is a constant less than 1, set to 0.95 here. 
$DR_{i}$ is the required draining rate to provide empty queue before new video frame comes.
The goal is to utilize 95\% of the available bandwidth, reserving some capacity to achieve lower latency.

$y^{+}$ represents the positive operator, meaning that if the rate required to drain the queue at the base station exceeds the allocated bandwidth, the feedback estimated available bandwidth value to the sender is set to 0.The base station attaches this rate value to the ACK packet header and transmits the information back to the sender.

\textit{\textbf{\underline{Challenge 3:}} Despite lacking prior knowledge of the flow’s actual frame size and frame interval, the base station must accurately estimate both the total data volume entering its queues before the sender updates the bitrate and the available queue drainage time, thus deriving a draining rate capable of timely queue clearance.}

Given: 
\begin{equation}
    DR_{i}= \frac{PredQ_{i}}{FI_{i}}
    \label{eq10bwnew_1}
\end{equation}
\noindent where $PredQ_{i}$ is the estimated queue length, $FI_{i}$ is the estimated frame interval.

We estimate frame interval $FI_{i}$ firstly. 
Video frames are generated periodically (e.g., 16.6 ms intervals for 60 fps streams), while packets within the same frame exhibit minimal inter-packet intervals.

Leveraging this property, two consecutively enqueued packets are classified into distinct frames if their inter-packet interval exceeds a predefined threshold. 
By identifying the tail packet timestamp of frame $n$ and the head packet timestamp of frame $n+1$, alongside the head packet timestamp of frame $n$, the frame interval can be calculated.

To enhance prediction accuracy, we apply an Exponential Weighted Moving Average (EWMA) to historical and current frame interval values.
\begin{equation}
 FI_{i} = \alpha \cdot FI_{n-1} + (1-\alpha) \cdot FI_{n}
   \label{eq10bwnew_1_2}
\end{equation}
\noindent where the weight $\alpha$ is 0.8.

Next, we predict enqueued and dequeued data volumes based on the propagation delay between the base station and the sender. 

The dequeued data volume is calculated based on the average allocated bandwidth per frame interval $\overline{BW_{i}}$. 
Converting the frame interval $FI_{i}$ into the number of TTIs yields
\begin{equation}
    \overline{BW_{i}}= \frac {\sum_{j=1}^{N_{TTI}} BW_{i}}{N_{TTI}}
    \label{eq10bwnew_1_2_3}
\end{equation}

The enqueued data volume is calculated as the sum of downlink in-flight data frames and newly generated data frames during feedback uplink transmissions.

The number of all in-flight data frames and newly generated data frames is:
\begin{equation}
    FNum_{i}=\lfloor\frac{ 2\times WiredND_{i}}{FI_{i}}\rfloor+1
    \label{eq10bwnew_2}
\end{equation}
The wired network delay \textbf{$WiredND_{i}$} of flow $i$ is measured using ping between the base station and the server.

\begin{equation}
    T_{i}= (TTI+FI_{i}  \cdot FNum_{i})
    \label{eq10bwnew_1_1}
\end{equation}
$T_{i}$ represents the time between the timestamp that current feedback reaches the sender and the timestamp that newly generated frame arrive at the base station.
This time is estimated based on all frames that were sent before the next bitrate update but have not yet arrived at the base station, also includes a TTI of one base station scheduling period.
\begin{equation}
    LowerT_{i}= NowT - FI_{i} \cdot (FNum_{i}-1) - 2\cdot WiredND_{i}
    \label{eq10bwnew_3}
\end{equation}
\begin{equation}
    UpperT_{i}= NowT - FI_{i} \cdot FNum_{i} - 2\cdot WiredND_{i}
    \label{eq10bwnew_4}
\end{equation}
$LowerT_{i}$ and $UpperT_{i}$ represent the lower and upper bounds of the calculated delay interval, respectively, while $NowT$ denotes the current time when the base station executes the algorithm.  

Using this delay interval, the most recently recorded feedback ACK at the base station can be retrieved, along with its corresponding feedback rate $HisBW_{k}$. 

Finally, we successfully capture the new base station queue length $PredQ_{i}$ of flow $i$ during the decision-to-execution latency between the base station and sender.

\begin{equation}
    PredQ_{i}= \sum_{k=1}^{FNum} (HisBW_{k}-\hat e_{k})\times  FI_{i}+rlcQ_{i}-\overline{BW_{i}} \times T_{i}
    \label{eq10bwnew_6}
\end{equation}

$\hat e_{k}$: An error correction term for $HisBW_{k}$ is derived from autoregressive statistical analysis of discrepancies between the historically predicted bitrate and the actual transmitted bitrate.

$rlcQ_{i}$: The base station system uses TTI as the smallest time granularity, and the queue length of the RLC layer buffer for user equipment n is sampled in real time during each TTI. 
To effectively characterize the actual utilization efficiency of channel resources, we employ a sliding time window-based minimum queue statistic, denoted as $RLC_n$.
This metric is defined as the minimum queue length among all sampled instances within the most recent frame interval. 

This design of $rlcQ_{i}$ is motivated by an in-depth analysis of the bursty transmission characteristics of video streaming. 
Traditional methods, such as moving averages or instantaneous queue indicators, suffer from inherent limitations. 
The former, due to its smoothing effect, fails to promptly capture bursty load variations, while the latter is susceptible to transient arrival fluctuations that may introduce noise into the information retrieval process. 
By extracting the minimum queue length within the time window, this approach effectively mitigates false peak interference caused by bursty traffic while accurately capturing the lower bound of the channel’s carrying capacity.

        

\begin{algorithm}
    \SetAlgoLined
    \KwData{$FI_{n-1}$, $FI_n$, $\alpha$, $BW_i$, $N_{TTI}$, $WiredND_i$, $TTI$, $NowT$, $HisBW_k$, $\hat e_k$, $rlcQ_i$, $\eta$}
    \KwResult{$\widehat{BW_i}$}
    \BlankLine
    \SetKwFunction{FMain}{Available Bandwidth Prediction}
    \SetKwProg{Fn}{function}{:}{}
    \Fn{\FMain{}}{
        $FI_i \leftarrow \alpha \cdot FI_{n-1} + (1-\alpha) \cdot FI_n$\;
        $\overline{BW_i} \leftarrow \frac{\sum_{j=1}^{N_{TTI}} BW_i}{N_{TTI}}$\;
        $FNum_i \leftarrow \left(\left\lfloor \frac{2 \times WiredND_i}{FI_i} \right\rfloor + 1\right)$\;
        $T_i \leftarrow TTI + FI_i \cdot FNum_i$\;
        $BW_n \leftarrow \sum_{i=1}^{N_t} BW_{ni}$\;
        $PredQ_i \leftarrow \sum_{k=1}^{FNum} (HisBW_k - \hat e_k) \times FI_i + rlcQ_i - \overline{BW_i} \times T_i$\;
        $DR_i \leftarrow \frac{PredQ_i}{FI_i}$\;
        $\widehat{BW_i} \leftarrow \left(\eta \times \overline{BW_i} - DR_i\right)^{+}$\;
        
        \KwRet{$\widehat{BW_i}$}
    }
    \caption{Flow Pattern-Oriented Available Bandwidth Prediction}
    \label{alg:flow-pattern-bandwidth-prediction}
\end{algorithm}

\subsection{User experience oriented rate control}
This module translates the base station's guidance bandwidth $\widehat{BW_{i}}$ into two executable parameters: a \textbf{target bitrate} for the sender encoder and a \textbf{pacing rate} for the congestion controller, preventing sender-side queue buildup. It further supports bitrate adaptation based on user-specific experience requirements.

\textbf{Impact of bitrate dynamics in the encoder.} There are performance differences between various encoders. Standard encoders may take 2-3 frames to adjust from the current bitrate to the target bitrate, with each adjustment having a small bitrate gradient. 
However, high-performance encoders can immediately reach the target bitrate within a single frame, which can cause frequent fluctuations in actual bitrate under Choir's mode. 

Based on the available bandwidth feedback from the base station, the sender calculates and reports the target bitrate to the encoder to ensure that the bitrate adapts to the available bandwidth. 
Additionally, it is necessary to ensure that the pacing rate of the sender buffer is sufficiently larger than the actual bitrate to avoid packet queuing at the sender.

The sender can directly use the $\widehat{BW}$ returned by the base station as the target bitrate passed to the encoder to achieve the lowest network transmission delay. 

However, since the $\widehat{BW}$, which takes into account the base station's queue draining rate, can fluctuate significantly, adjusting the encoder's bitrate in real-time based on this value may lead to visual quality fluctuations that could affect the user experience. 

Therefore, the sender can smooth this value over a period of time. 
This module introduces a smoothing coefficient $\epsilon$, which is currently set to 1, meaning that the target bitrate $BR_{i}$ adopts current allocated bandwidth $\widehat{BW}$ directly. 

$\epsilon$ set to 10 means that the historical value of the target bitrate assigned to the encoder for the last 9 times and the current $\widehat{BW}$ will be smoothed. 

Thus, the target bitrate $BR_{i}$ to be reported to the encoder at decision moment is given by:
\begin{equation}
    BR_{i}=\frac{\sum_{i=1}^{\epsilon-1} \cdot HisBR_{n} +\widehat{BW_{i}}}{\epsilon}
    \label{eq13}
\end{equation}
\noindent where $HisBR_{n}$ represents the historical bitrate record of each frame $n$ generated by the sender's encoder.

And the pacing rate of congestion controller is:
\begin{equation}
    PR_{i}=\rho \cdot max(ReceiveRate, BR_{i})
    \label{eq14}
\end{equation}

The sender calculates the receiving rate based on the acknowledged byte count from ACK packets and selects the maximum receiving rate ($ReceiveRate$) over a past time window as a reference. 
The pacing rate must be greater than both $ReceiveRate$ and the upcoming bitrate assignment $BR_{i}$. 
To ensure this, we set the scaling factor $\rho$ empirically to 1.25 in our experiments.

By employing a variable pacing rate, we ensure that the sender's transmission rate remains sufficient, while preventing the send buffer from becoming the primary bottleneck that increases latency. 
This approach shifts congestion primarily to the wireless link bottleneck, allowing more effective bandwidth adaptation.

\section{Implementation}
We first configured a simulation environment to import highly fluctuating traces for single-stream performance testing. 
A RAN testbed with a capacity of 30Mbps was used to validate simulation results. We deployed a RAN testbed with a capacity of around 300Mbps to support concurrent experiments with multiple high-bitrate streams.

\subsection{Trace-driven 5G Module Simulation}
Considering 5G emulation limitations on Mahimahi\cite{sentosacellreplay}, we adopt the 5G LENA module of the NS3 to explicitly configure TDD patterns, achieving realistic dynamic delay\cite{patriciello2019e2e, lecci2021ns}. 
The 5G LENA is a new radio (NR) network module and incorporates physical layer, Medium Access Control (MAC) layer, and RLC layer features aligned with NR Release 15 TS 38.300\cite{lena}. 

Considering the potential variability of wired network latency between senders and 5G base stations in real-world scenarios, we set sender-to-base station latency to 1ms, 10ms, and 20ms in simulations to observe performance of different solutions across conditions.

\subsection{Software-defined 5G Open-RAN Testbed}
\begin{figure}[tbp]
    \centering
    \includegraphics[width=1\linewidth]{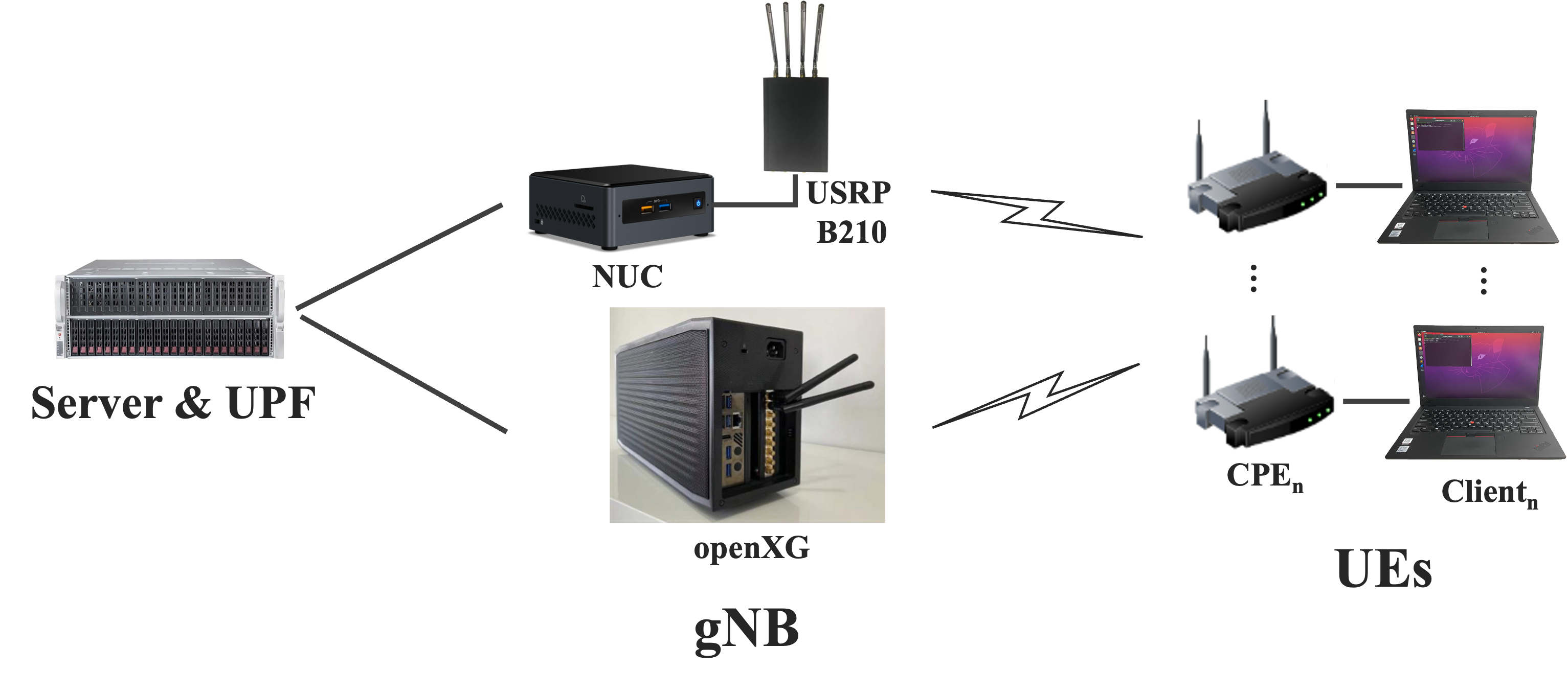}
    \caption{Open-source 5G network testbed} 
    \label{fig:34}    
\end{figure}

As shown in Figure~\ref{fig:34}, we establish two base station systems with different capacities. 
The upper system uses a host equipped with the open-source OAI 5G NR protocol stack\cite{kaltenberger2020openairinterface} and a USRP B210 as the radio frequency board. 
The bottom system integrates the OAI 5G NR protocol stack with the high-performance radio frequency board of the Witcomm OpenXG\cite{xgproduct}.

The basic framework of the deployed OAI 5G NR protocol stack complies with the requirements of 3GPP R15, with upgrades referencing key features from 3GPP R16 and R17, including the Radio Resource Control (RRC) protocol and the Radio Link Control (RLC) protocol\cite{openairinterface5g, TS38331, TS38322}. 
Both systems support the Hybrid Automatic Repeat reQuest (HARQ) mechanism at the MAC layer and the Acknowledgement Mode (AM) retransmission mechanism at the RLC layer\cite{3gpp.23.288}.

In our network topology, the open-source core network free5GC is deployed on a standalone server, with the core network and applications isolated using Docker. 
These components are connected to both base stations via wired links, with a physical bandwidth of 1Gbps.

We use a combination of Customer-Premises Equipment (CPE) and laptops as receiver to facilitate more concurrent streams and data collection. 
The OAI 5G UE protocol is deployed on the laptops.

\begin{table}
    \centering
  \caption{RAN Configuration}
  \label{table:31q}
  \begin{tabular}{ccc}
    \toprule
    \textbf{Parameters}&\textbf{ in B210 }&\textbf{ in openXG }\\
    \hline
    Bandwidth& 40 MHz&100 MHz\\
    \hline
    Frequency & \multicolumn{2}{c}{3.5 GHz}\\
    \hline 
    TDD Pattern & \multicolumn{2}{c}{DDDSU}\\
    \hline
    Sub-Carrier Space & \multicolumn{2}{c}{30 KHz}\\
  \bottomrule
\end{tabular}
\end{table}

\subsection{Upper Layer Protocols}
The QUIC is deployed as the transport layer protocol for both simulation and testbed. We adopt the Ali XQUIC, which supports datagram mode transmission\cite{xquic3}.

For the application layer, we refer to the draft specification of RTP over QUIC for custom development in this environment\cite{ietf-avtcore-rtp-over-quic-11}. 
In terms of congestion control mechanisms, the RTCP feedback is removed, and instead, XQUIC’s ACK feedback is used to provide network status information to the sender. 
The transport layer's Congestion Control Algorithm (CCA) is responsible for adjusting the pacing rate or congestion window.

Additionally, the gstreamer encoder, which supports H.264, is integrated with XQUIC for inter-process communication. 

\subsection{Feedback to Sender}
To reduce feedback latency and enhance the efficiency and timeliness of feedback information, this approach adopts an online feedback mechanism. 
The base station directly marks feedback information within uplink data packets, thereby shortening the feedback path.

Given the complexity of QUIC's encrypted payload mechanism, Choir utilizes the IP option field to carry base station feedback information.
Specifically, a new field is designed within the IP option to encapsulate the feedback rate information from the base station. 
The implementation must support encoding both the rate value and the corresponding unit length within the option field. \cite{li-6man-apn-ipv6-encap-00}.

\section{Evaluation}
\subsection{Experiment Setup}
First, all encoded video streams have a fixed frame rate of 60FPS, so the theoretical video frame interval is 16.6ms.

\textbf{Baselines.} The baselines include both end-side solutions and network-assisted solutions. 

For end-side solutions, we deploy:
\begin{itemize}
    \item Pudica \cite{wang2024pudica}: A flow control solution that precisely controls delay using bandwidth utilization rate, integrated with XQUIC.
    \item SQP \cite{ray2022sqp}: A delay-based congestion control algorithm that probes bandwidth using precise sending control of each frame, integrated with XQUIC.
    \item SCReAM \cite{johansson2017scream}: A widely used delay-based rate control algorithm.
    \item COPA \cite{arun2018copa}: A delay-based congestion control algorithm designed for wireless networks, implemented with $\delta$ = 0.1.
\end{itemize}

Here, we did not include additional comparisons with more bitrate control algorithms such as GCC\cite{carlucci2016analysis}, and NADA\cite{zhu2020NADA}, as previous research shows SCReAM has better latency performance compare to GCC and NADA\cite{zhang2019congestion}.  

For network-assisted solutions, we deploy:
\begin{itemize}
    \item L4S\cite{l4s}: A feedback-driven collaboration solution where the base station predicts congestion variations by sensing queueing delay thresholds and informs the sender of "increase" or "decrease" bandwidth signals through the receiver.
    \item ABC\cite{goyal2020abc}: A solution where network devices such as access points actively provide "acceleration" and "deceleration" signals by accurately predicting the forwarding bandwidth.
    \item SCONE\cite{shi-scone-rtc-requirement-02}: A collaboration solution where the base station informs the sender of bandwidth capacity and queue length, implemented with draining time as 16.6ms.
\end{itemize}

PBE-CC is excluded from the baseline since it only implements bandwidth feedback without addressing queueing impacts for video streams. Theoretically, even with queue drainage function, PBE-CC underperforms SCONE due to delayed feedback from receiver-side.

\textbf{Traces.} The simulation environment evaluation was conducted using traces collected from real 5G dynamic network. 

\textbf{Metrics.} We use the following metrics for evaluation.
\begin{itemize}
    \item Frame delay. The time between frame encoding and decoding, reflecting real-time latency performance. We measure average delay, 95th percentile tail delay, and 99.9th percentile tail ldelay.
    \item Average Bitrate. The common indicator of video quality, with larger rate indicating higher visual fidelity.
\end{itemize}

\subsection{Performance of single flow transmission}

We varied the wired network latency between the sender and the base station to investigate the impact of different latency on algorithm effectiveness. 

\begin{figure*}[htbp]
    \centering
    \begin{subfigure}[b]{0.24\textwidth}
        \centering
        \includegraphics[width=\textwidth]{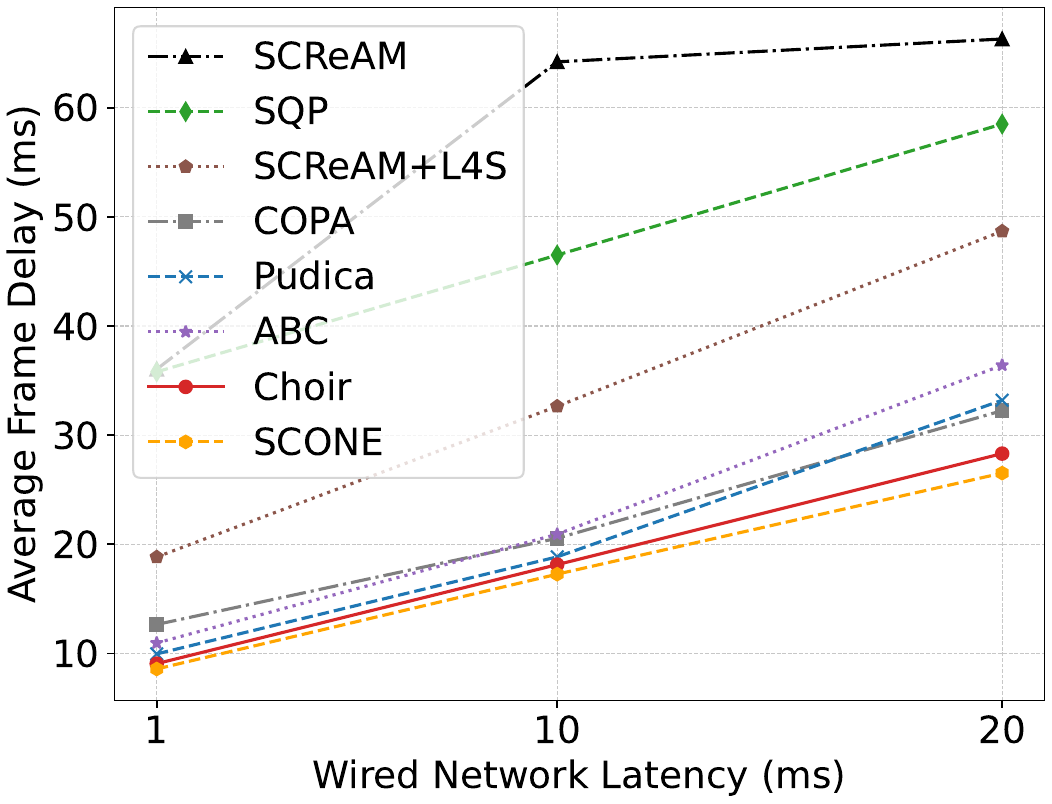}
        \caption{Average Frame Delay}
        \label{fig4:3}
    \end{subfigure}
    \begin{subfigure}[b]{0.24\textwidth}
        \centering
        \includegraphics[width=\textwidth]{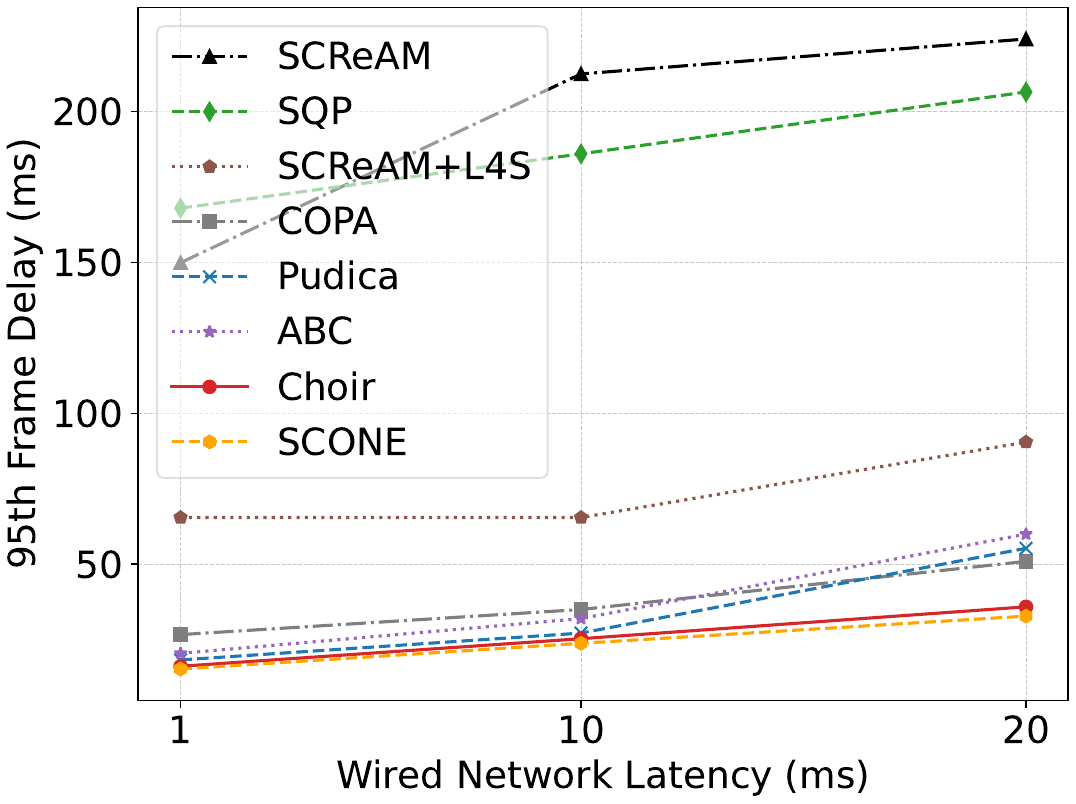}
        \caption{95th Frame Delay}
        \label{fig4:1}
    \end{subfigure}
    \begin{subfigure}[b]{0.24\textwidth}
       \centering
       \includegraphics[width=\textwidth]{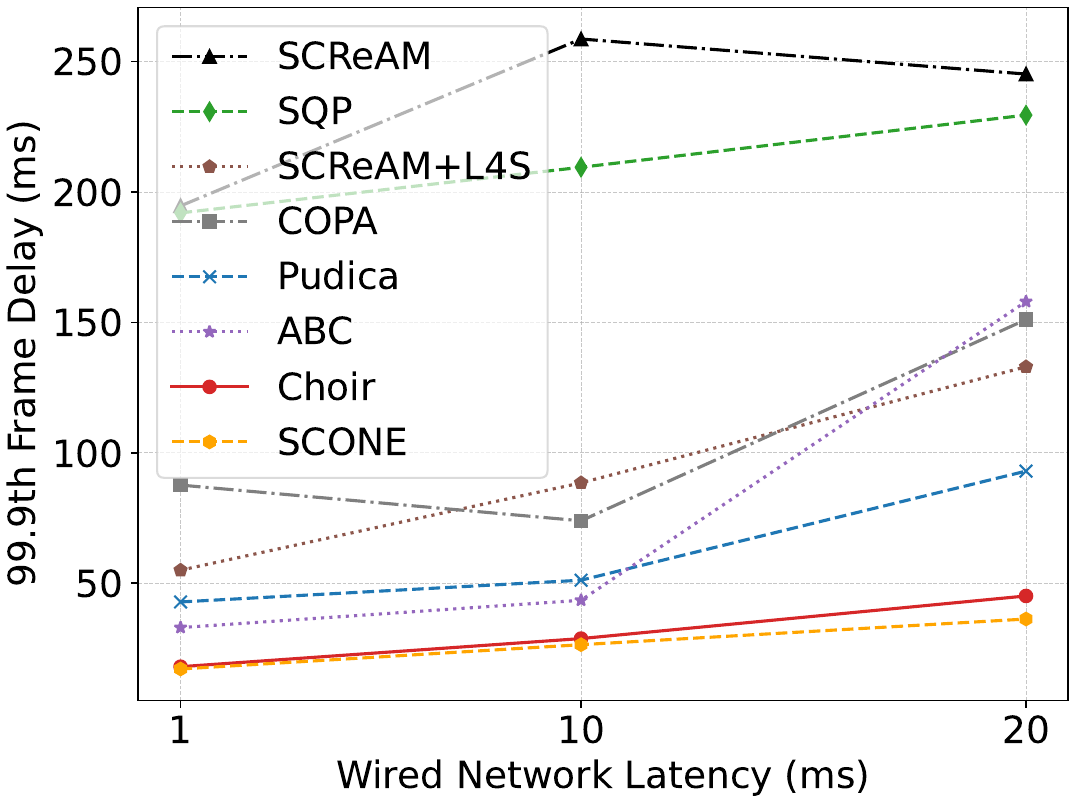}
       \caption{99.9th Frame Delay}
       \label{fig4:2}
    \end{subfigure}
    \begin{subfigure}[b]{0.24\textwidth}
        \centering
        \includegraphics[width=\textwidth]{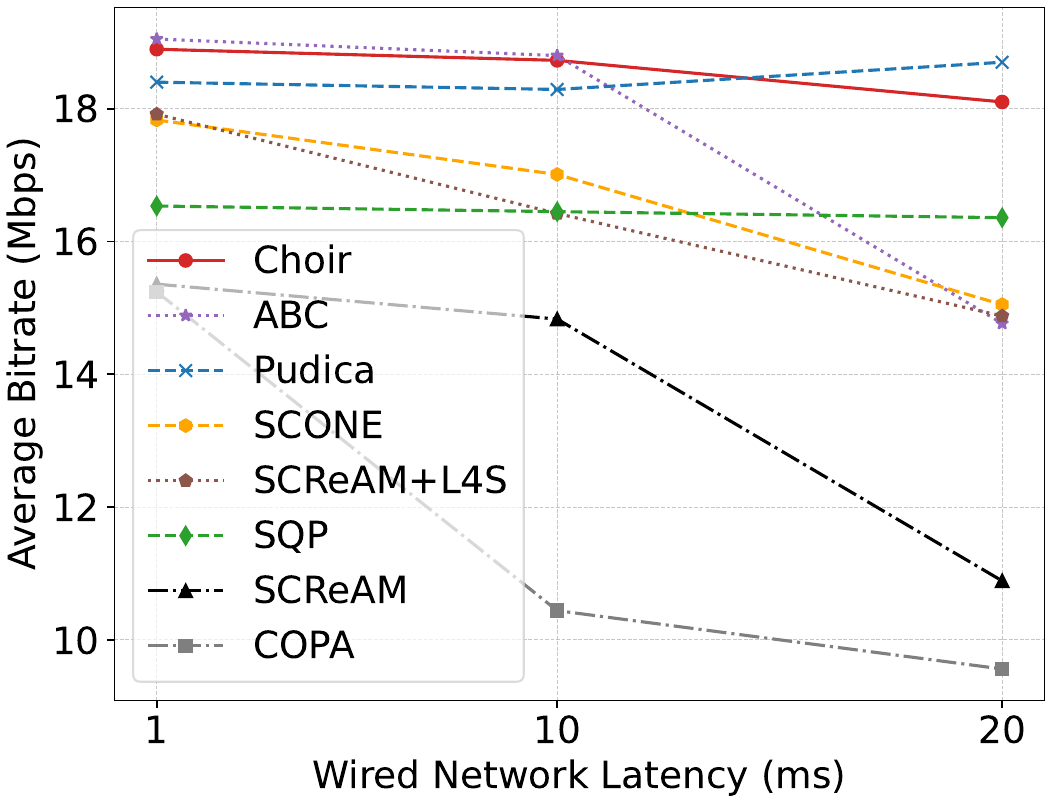}
        \caption{Average Bitrate}
        \label{fig4:4}
    \end{subfigure}   
    \caption{Bitrate and frame delay achieved by 8 control solution in trace-driven simulation.}
    \label{fig:trace-driven simulation}
\end{figure*}

As shown in Fig.~\ref{fig:trace-driven simulation}, Choir outperforms other solutions in the frame delay and the average bitrate. 
With increasing wired network latency, both end-side solutions and network-assisted solutions suffer from delayed network state reactions, leading to uncontrollable increases in frame tail delay or decreases in average bitrate.

For 99.9th percentile tail delay: Under 1 ms wired network latency, while SCONE achieves 4.9\% lower tail latency than Choir, Choir still reduces latency by 45.5\% to 90.7\% compared to other solutions. In 10 ms wired network latency, SCONE maintains an 8.1\% latency advantage over Choir; however, Choir exhibits a 33.8\% to 88.9\% latency reduction relative to other solutions. Even under 20 ms wired network latency, where SCONE outperforms Choir by 19.2\% in tail latency, Choir sustains a 51.4\% to 81.6\% latency reduction compared to other solutions. 

In terms of average bitrate, Choir exhibits minimal trade-offs against top-performing solutions while significantly outperforming lower-performing alternatives across varied network latency conditions. 
Under 1 ms wired network latency, Choir reduces the average bitrate by a marginal 0.7\% compared to the best-performing method but exceeds lower-performing solutions by 2.6\% to 19.3\%. 
At 10 ms wired network latency, the reduction relative to the top method narrows to 0.4\%, while improvements over lower-performing solutions range from 2.3\% to 79.5\%. 
Under 20 ms wired network latency, Choir’s average bitrate decreases by 3.3\% compared to the optimal baseline but achieves an 11.0\% to 89.9\% enhancement over suboptimal alternatives.
\begin{figure*}[htbp]
    \centering
    \begin{subfigure}[b]{0.24\textwidth}
        \centering
        \includegraphics[width=\textwidth]{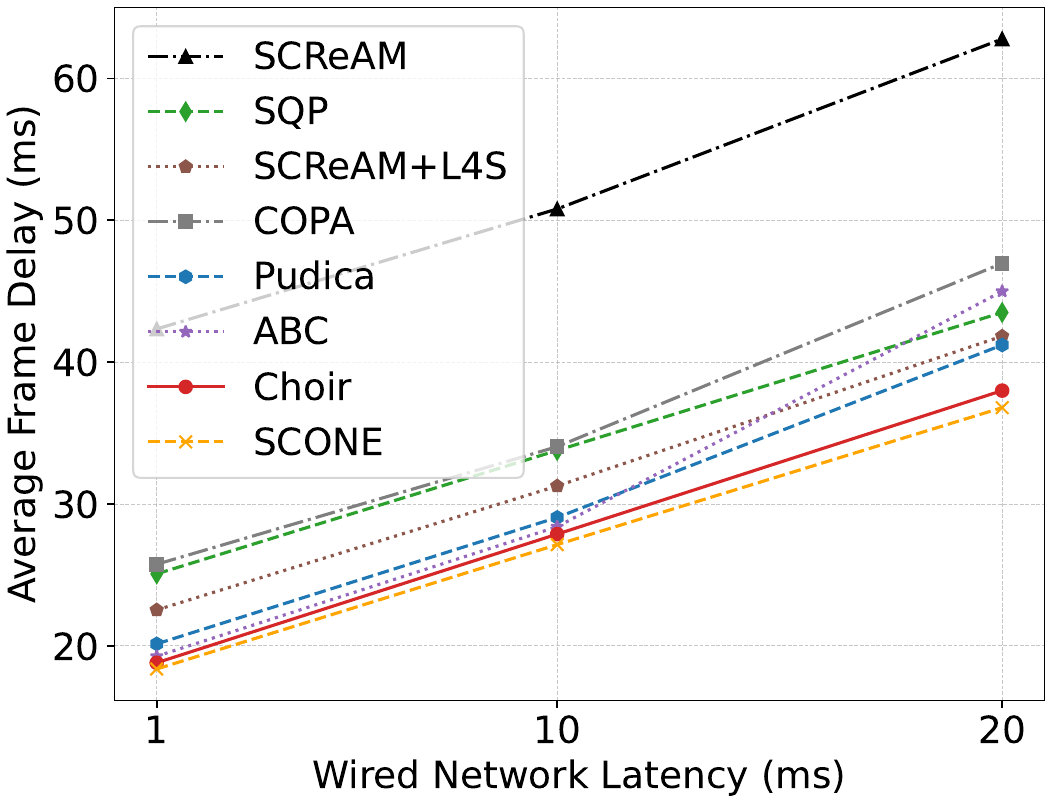}
        \caption{Average Frame Delay}
        \label{fig4:3}
    \end{subfigure}
    \begin{subfigure}[b]{0.24\textwidth}
        \centering
        \includegraphics[width=\textwidth]{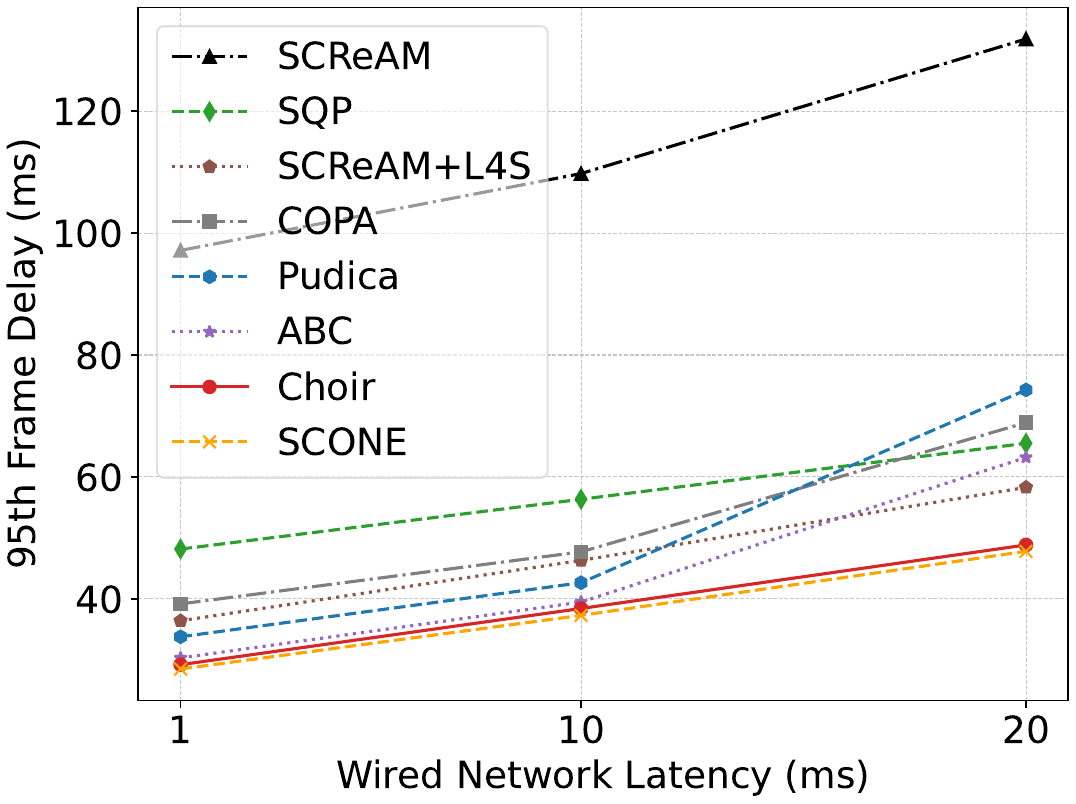}
        \caption{95th Frame Delay}
        \label{fig4:1}
    \end{subfigure}
    \begin{subfigure}[b]{0.24\textwidth}
       \centering
       \includegraphics[width=\textwidth]{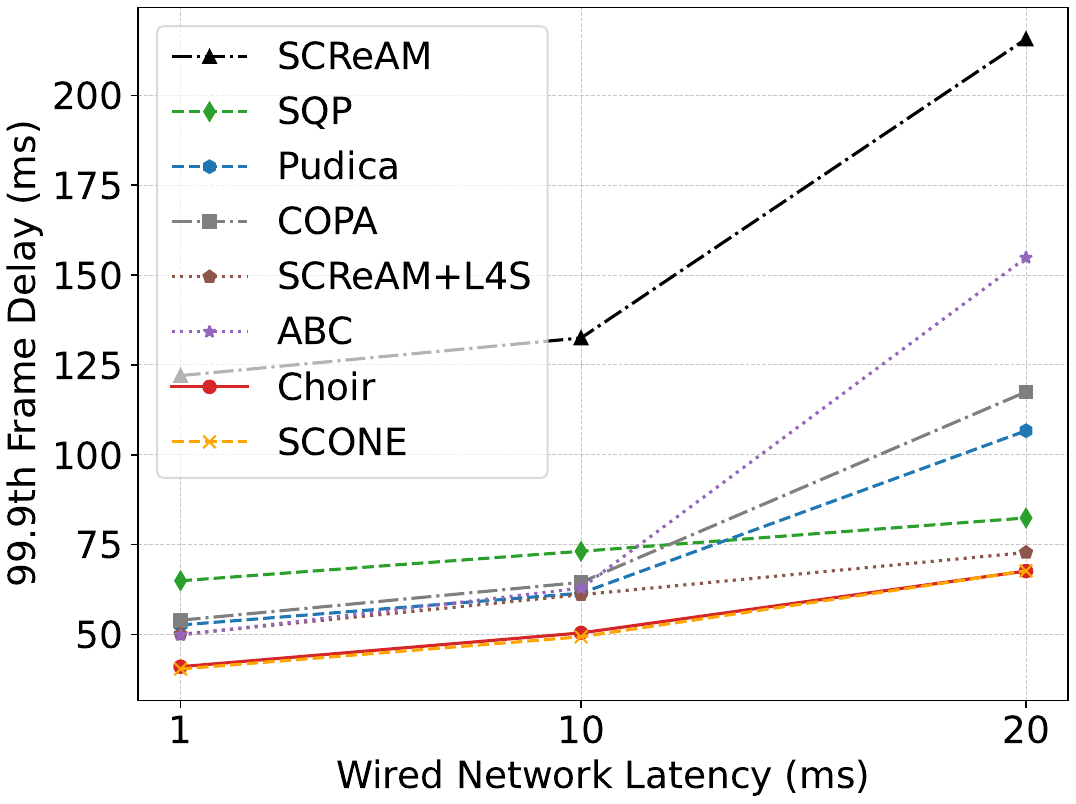}
       \caption{99.9th Frame Delay}
       \label{fig4:2}
    \end{subfigure}
    \begin{subfigure}[b]{0.24\textwidth}
        \centering
        \includegraphics[width=\textwidth]{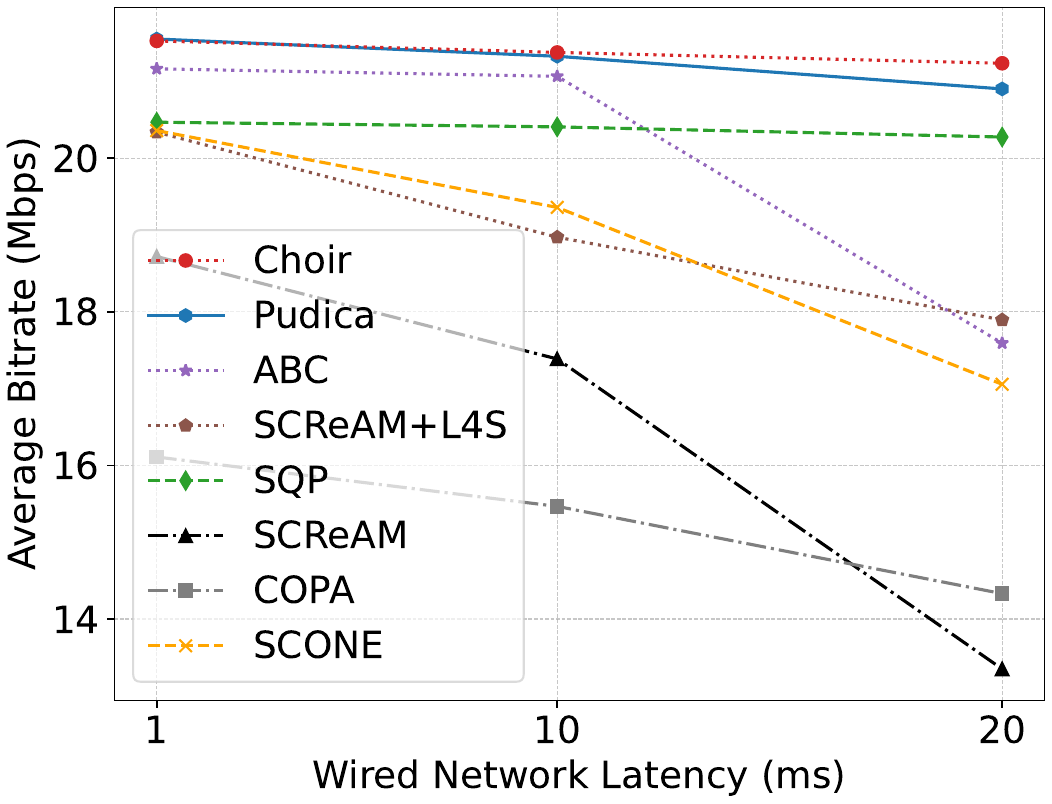}
        \caption{Average Bitrate}
        \label{fig4:4}
    \end{subfigure}   
    \caption{Bitrate and frame delay achieved by 8 control solution in B210 RAN.}
    \label{fig:B210 RAN}
\end{figure*}

Compared to other solutions, Choir leverages its predictive rate control design and demonstrates superior performance at higher wired network latency while maintaining high frame quality. 

Fig.~\ref{fig:B210 RAN} further proves Choir achieves the nearly lowest frame delay while maintaining the highest average bitrate, among the 8 solutions tested in the B210 RAN. Choir also ensures minimal variation in average bitrate.

For 99.9th percentile tail latency, in 20 ms network latency scenarios, SCONE’s tail latency becomes comparable to Choir, while Choir maintains 7.10\% to 68.64\% latency reduction over existing solutions.
\begin{figure}[htbp] %
        \centering
        \begin{subfigure}[b]{0.49\linewidth}
            \centering
            \includegraphics[width=\textwidth]{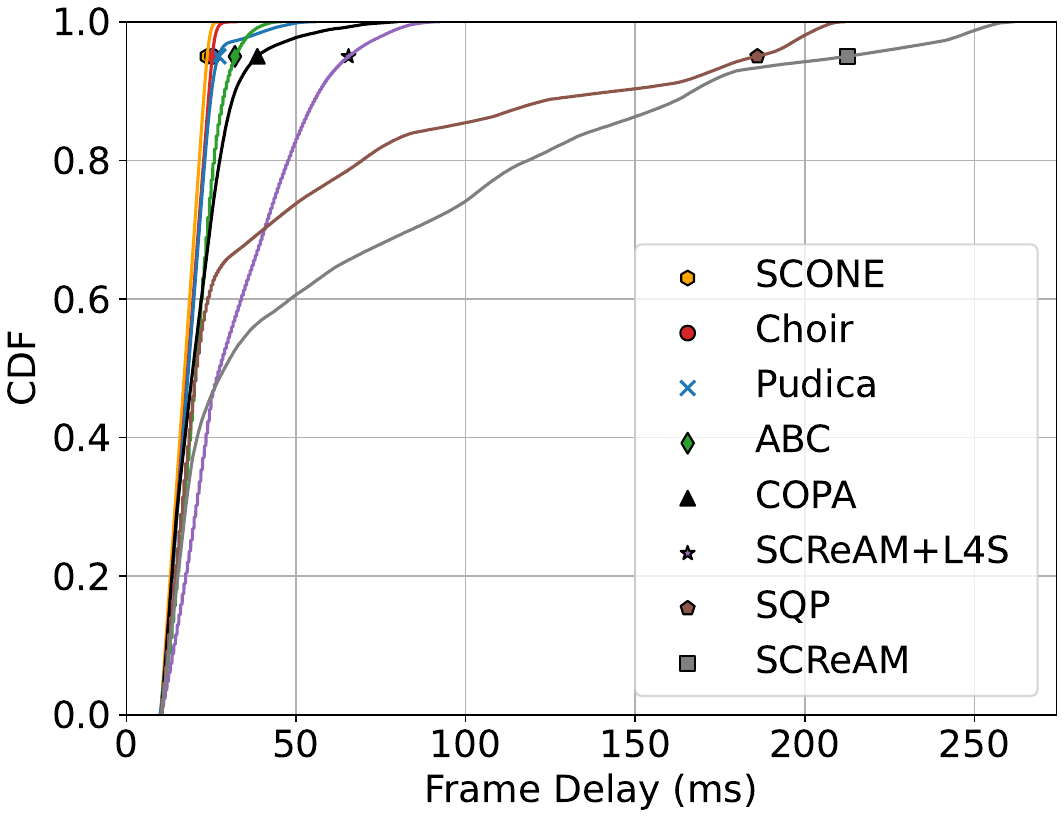} %
            \caption{Delay}
            \label{fig:delay_cdf_10ms}
        \end{subfigure}
        \begin{subfigure}[b]{0.49\linewidth}
            \centering
            \includegraphics[width=\textwidth]{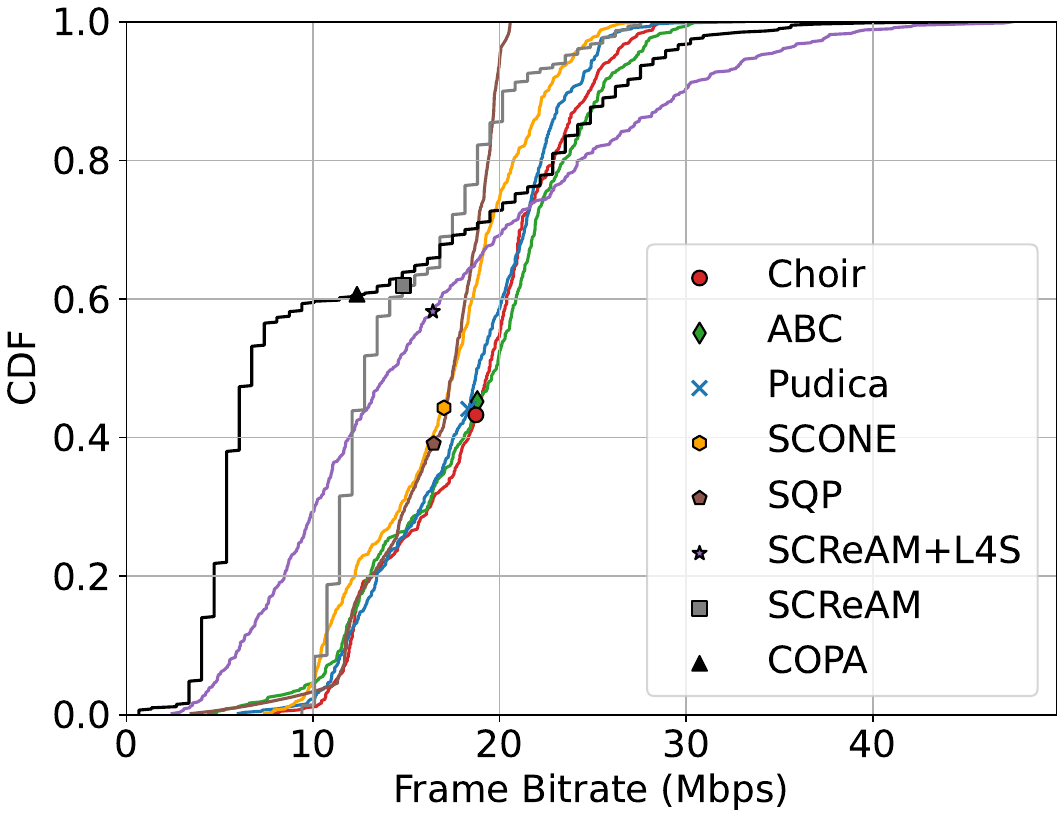} %
            \caption{Bitrate}
            \label{fig:bitrate_cdf_10ms}
        \end{subfigure}
        \caption{The distribution of delay (a) and bitrate (b) of 8 control solution in 10ms wired network latency.}
        \label{fig:delay_bitrate_cdf}
\end{figure}
In terms of average bitrate, at 10 ms wired network latency, Choir surpasses the SOTA baseline by 0.23\% and outperforms weaker solutions by 1.45\% to 27.64\%. Under 20 ms wired network latency, Choir achieves a 1.57\% improvement over the optimal benchmark while delivering 4.52\% to 37.13\% higher bitrates than other baselines.

When the wired network latency exceeds the update interval of the encoder, the returned available bandwidth will lag behind a bitrate adjustment, which is detrimental to the sender to react to instantaneous RAN fluctuations, potentially causing some queuing in the base station and increased latency.

The results of Choir show that the frame tail delay increase was nearly identical to the basic RTT growth, indicating that the impact of a longer feedback path on Choir is limited.
\begin{figure*}[htbp]
    \centering
    \begin{subfigure}[b]{0.24\textwidth}
        \centering
        \includegraphics[width=\textwidth]{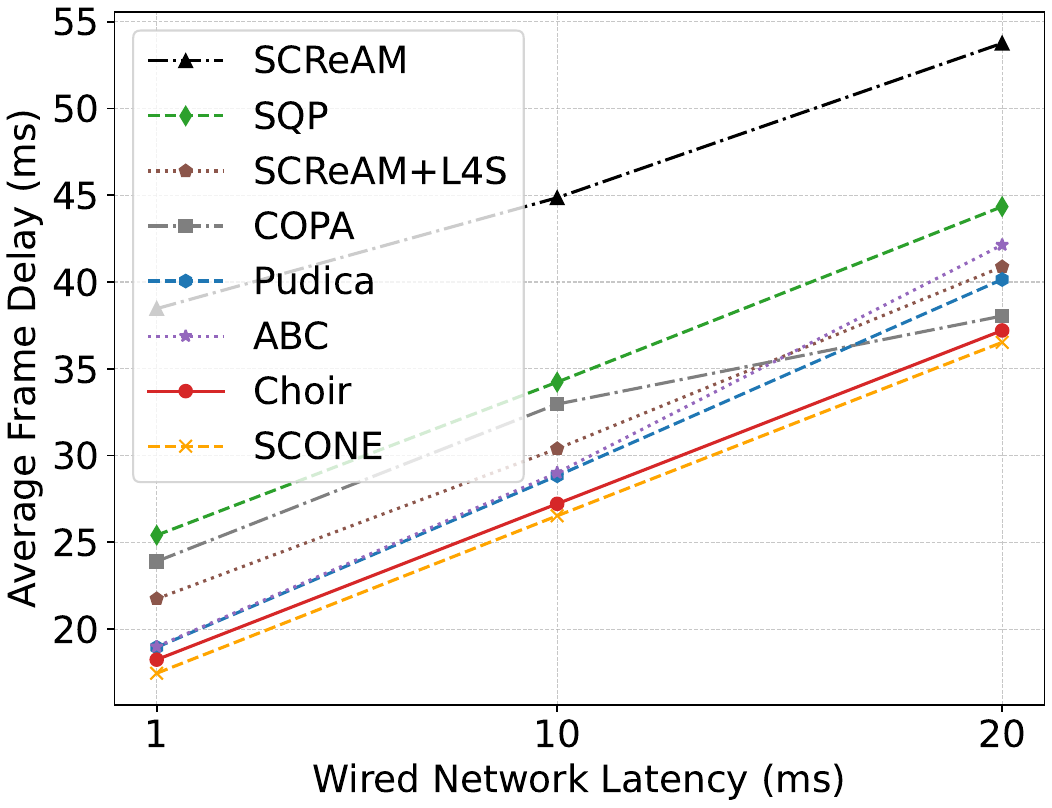}
        \caption{Average Frame Delay}
        \label{fig4:3}
    \end{subfigure}
    \begin{subfigure}[b]{0.24\textwidth}
        \centering
        \includegraphics[width=\textwidth]{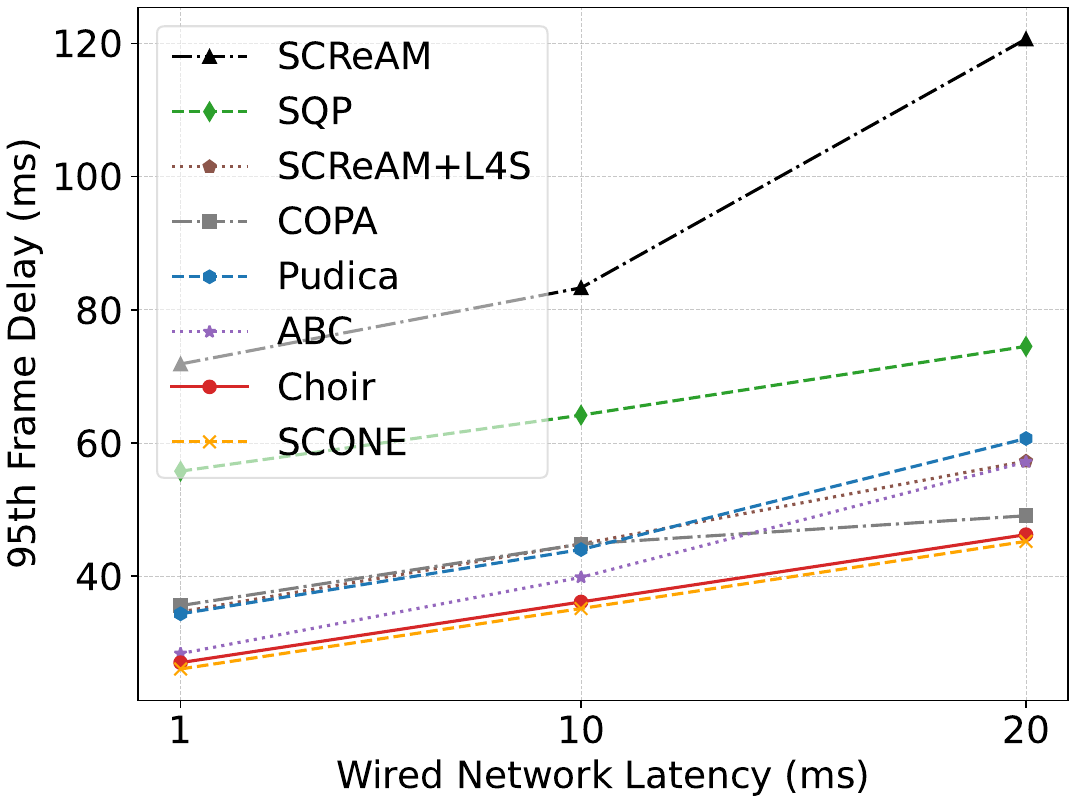}
        \caption{95th Frame Delay}
        \label{fig4:1}
    \end{subfigure}
    \begin{subfigure}[b]{0.24\textwidth}
       \centering
       \includegraphics[width=\textwidth]{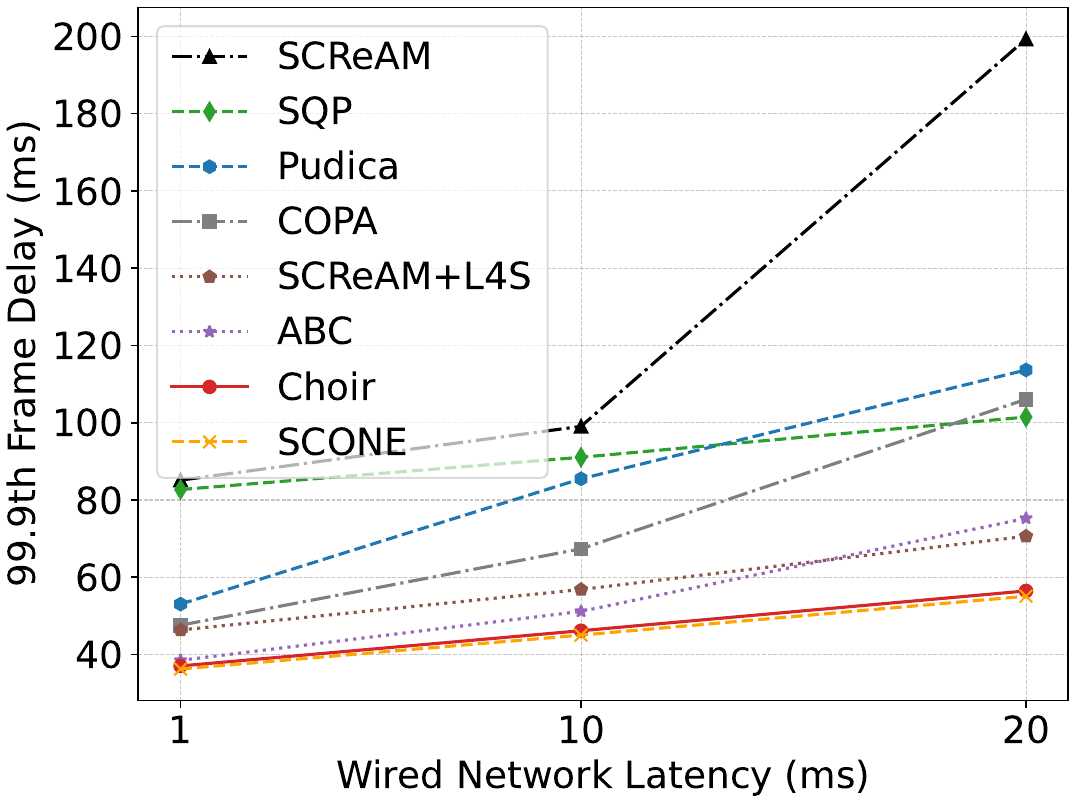}
       \caption{99.9th Frame Delay}
       \label{fig4:2}
    \end{subfigure}
    \begin{subfigure}[b]{0.24\textwidth}
        \centering
        \includegraphics[width=\textwidth]{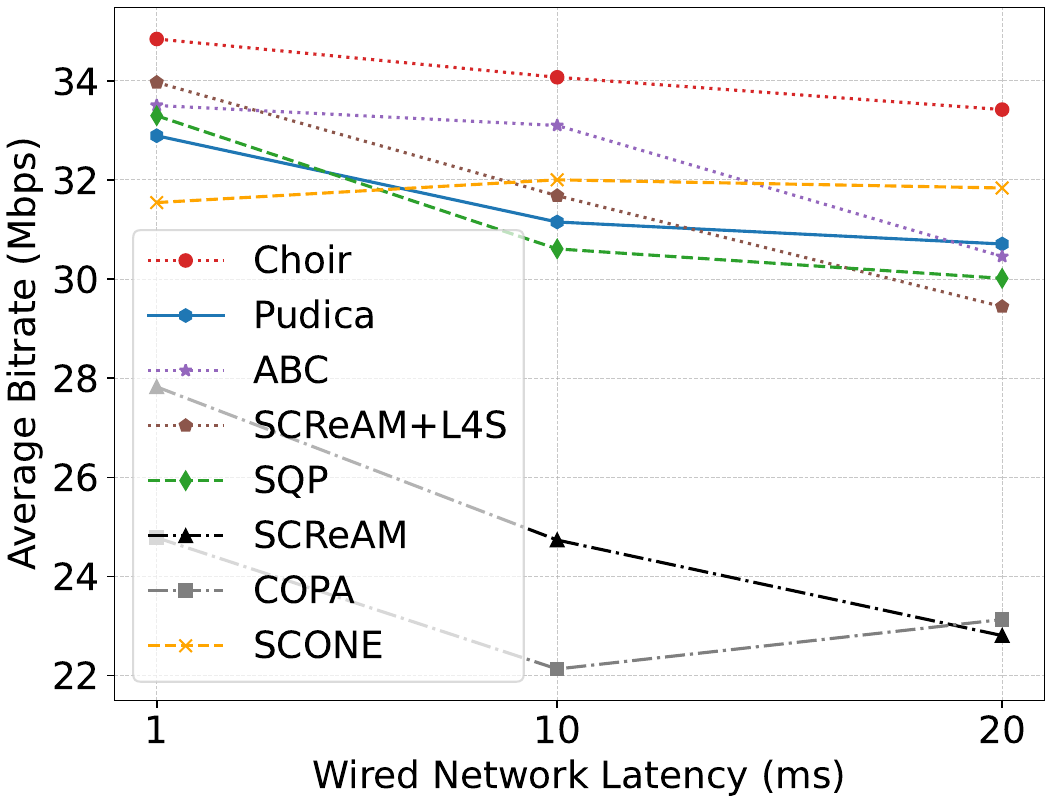}
        \caption{Average Bitrate}
        \label{fig4:4}
    \end{subfigure}   
    \caption{Bitrate and frame delay achieved by 8 control solution in openXG RAN.}
    \label{fig:openxg RAN}
\end{figure*}
Overall, Choir’s holistic performance surpasses SOTA rate control solutions, striking an comprehensive optimal performance between latency, throughput, and fairness.
\begin{figure}[! htbp]
    \centering
    \vspace{-0.3cm} %
    \begin{minipage}{\linewidth}
        \centering
        \setlength{\abovecaptionskip}{0.cm}
        \includegraphics[width=0.9\linewidth]{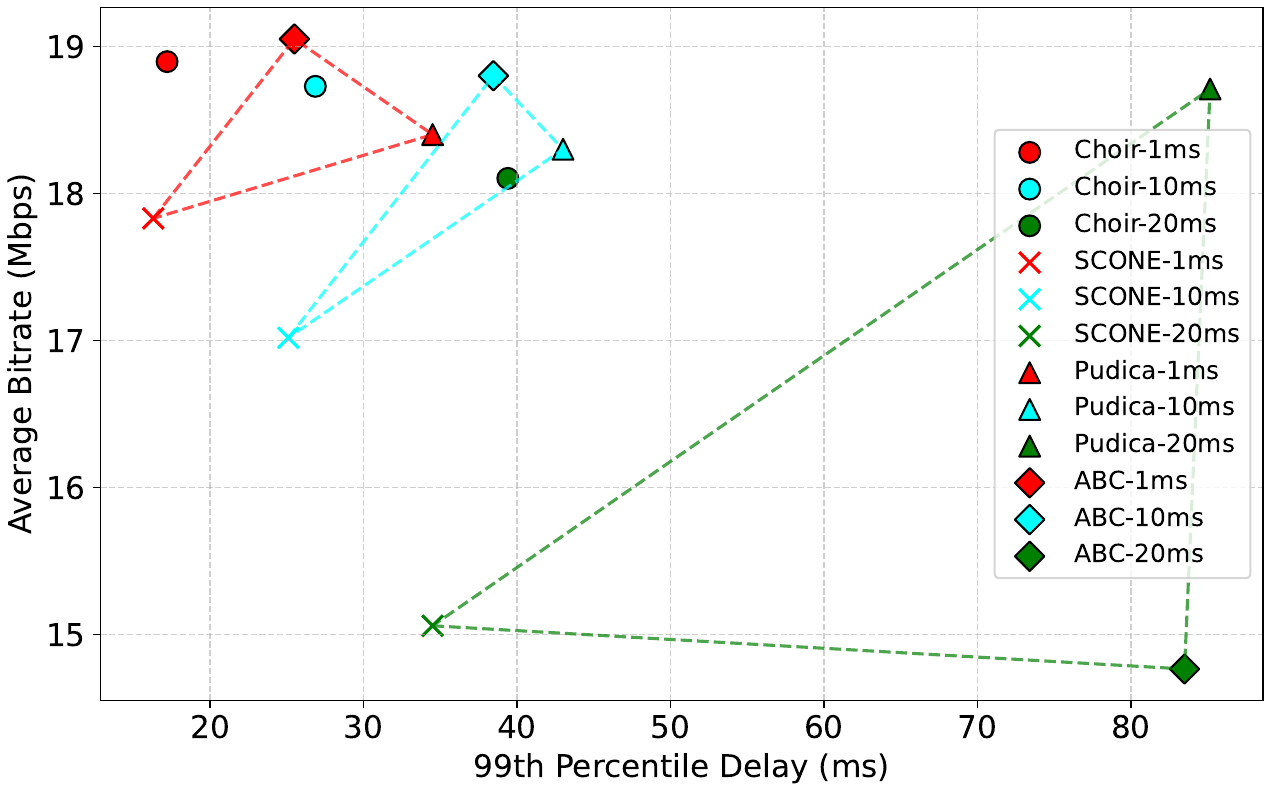}
        \caption{Comparison of bitrate and tail delay of 4 solutions.}
        \label{fig:fdcc_1_10_20ms_duibi}
    \end{minipage}   
    \vspace{-0.5cm} %
\end{figure}
As shown in Fig.~\ref{fig:delay_cdf_10ms}, markers are placed at the 95th percentile delay.

As shown in Fig.~\ref{fig:bitrate_cdf_10ms}, where markers are placed at the average bitrate, Choir maintains an equivalent average bitrate compared to higher-performing solutions, with only a 0.5\% reduction compared to the best-performing method. 
Compared to lower-performing solutions, Choir increases the average bitrate by 27\% to 80\%. These results demonstrate that Choir achieves lower video frame delay while maintaining higher bandwidth utilization.

As shown in Fig.~\ref{fig:fdcc_1_10_20ms_duibi}, we selected the best-performing solutions from the previous analysis: ABC, Pudica, and SCONE.
SCONE detects real-time queue buildup and forces the sender to drain it immediately, enabling extremely low tail delay.
But Choir consistently appears in the upper-left, indicating that it effectively balances frame delay and bandwidth utilization.

Compared to the baselines, Choir demonstrates faster and more stable adaptation to both bandwidth decreases and increases. 
When bandwidth suddenly decreases, Choir’s sender quickly responds to congestion, allowing it to rapidly drain the base station queue. 
This process is designed to be completed within a single frame interval in the optimal case. 
Once the queue is cleared, Choir restores its bitrate, ensuring efficient link utilization.

\subsection{Performance of multi-flows concurrent transmission}

We tested 7-flow concurrency schemes for 8 solutions to fully utilize the OpenXG RAN capacity. 
In the openXG RAN, the bitrate of each flow increased significantly. 
Larger frame sizes in high-bitrate video streams make it easier to cause severe congestion at the base station during multi-stream burst concurrency. 

Experimental results in the Fig.~\ref{fig:openxg RAN} demonstrate the performance of one flow in each 7-flow concurrency scenario. 
All Choir flows achieve fair resource utilization, resulting in consistent performance across flows. 
Other solutions exhibit uneven resource allocation, failing to ensure strict fairness. 
We select and show one flow with average performance.

\begin{table}
  \centering
  \caption{Performance of concurrent Choir flows at 1ms wired network delay}
  \label{table:openxgflows_1ms}
  \begin{tabular}{ccc}
    \toprule
    \textbf{Metrics} & \textbf{7 flow} & \textbf{14 flows} \\
    \hline
    P99.9 Delay (ms)   & 37.06           & 37.72           \\
    \hline
    P95 Delay (ms)   & 27.03           & 26.88           \\
    \hline
    Avg Delay (ms)   & 18.23           & 18.03           \\
    \hline
    Avg Bitrate (Mbps)   & 7 $\times$ 34.85  & 14 $\times$ 17.41\\
    \bottomrule
  \end{tabular}
\end{table}

\begin{table}
  \centering
  \caption{Performance of concurrent Choir flows at 10ms wired network delay}
  \label{table:openxgflows_10ms}
  \begin{tabular}{ccc}
    \toprule
    \textbf{Metrics} & \textbf{7 flow} & \textbf{14 flows} \\
    \hline
    P99.9 Delay (ms)   & 46.17           & 50.40           \\
    \hline
    P95 Delay (ms)   & 36.15           & 36.04           \\
    \hline
    Avg Delay (ms)   & 27.21           & 27.05           \\
    \hline
    Avg Bitrate (Mbps)   & 7 $\times$ 34.07  & 14 $\times$ 17.04\\
    \bottomrule
  \end{tabular}
\end{table}

For 99.9th percentile tail latency, under 1ms wired network latency, Choir reduces tail latency by 3.77\% to 56.48\% compared to other algorithms, except for SCONE, which exhibits 2.15\% lower tail latency than Choir. At 10ms wired network latency, Choir achieves 9.79\% to 53.41\% lower tail latency relative to alternative approaches, with SCONE marginally outperforming Choir by 2.45\%. Under 20ms wired network latency, Choir demonstrates significant latency reduction, ranging from 20.04\% to 71.67\% compared to other baselines, while SCONE maintains a slight advantage of 2.41\% lower tail latency over Choir.

For average bitrate, under 1ms wired network latency, Choir achieves an average bitrate improvement of 2.51\% to 28.88\% compared to existing solutions. At 10ms wired network latency, Choir demonstrates a more pronounced enhancement, increasing the average bitrate by 2.86\% to 35.04\% over alternative approaches. For scenarios with 20ms wired network latency, Choir maintains robust performance, delivering a 4.73\% to 31.76\% improvement in average bitrate relative to all baselines.

\begin{table}
  \centering
  \caption{Performance of concurrent Choir flows at 20ms wired network delay}
  \label{table:openxgflows_20ms}
  \begin{tabular}{ccc}
    \toprule
    \textbf{Metrics} & \textbf{7 flow} & \textbf{14 flows} \\
    \hline
    P99.9 Delay (ms)   & 56.47           & 74.258           \\
    \hline
    P95 Delay (ms)   & 46.25           & 46.23           \\
    \hline
    Avg Delay (ms)   & 37.20           & 37.10           \\
    \hline
    Avg Bitrate (Mbps)   & 7 $\times$ 33.42  & 14 $\times$ 16.70\\
    \bottomrule
  \end{tabular}%
\end{table}

Further, we scaled Choir’s concurrent flows from 7 to 14 to observe its performance under full base station load. 
As shown in Table ~\ref{table:openxgflows_1ms}, Table ~\ref{table:openxgflows_10ms} and Table ~\ref{table:openxgflows_20ms}, Choir maintains fair bandwidth allocation across 14 flows with no significant rise in tail delay.

\begin{figure}[tbp]
    \centering
    \includegraphics[width=0.8\linewidth]{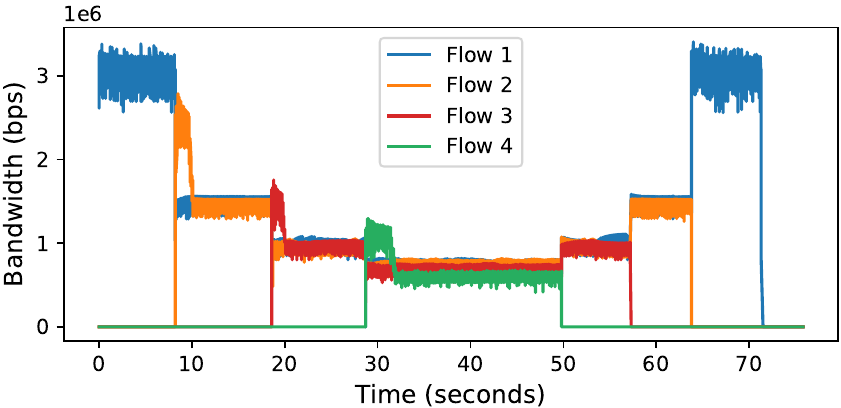}    
    \caption{Fairness of Choir.}
    \label{fig:fairness}
    \vspace{-0.5cm} %
\end{figure}

\textbf{Fairness.}The base station continuously collects, reclaims, and equally distributes bandwidth resources to all users with active queues. 
Therefore, whether it is fairness among multiple Choir flows, fairness between Choir flows and other flows, or background flows, the base station takes responsibility. Fig.~\ref{fig:fairness} shows the fairness performance of Choir during 4-stream transmission.

\subsection{Choir Deep Dive}

\begin{figure}[htbp]
    \centering
    \begin{minipage}{\linewidth}
        \centering
        \setlength{\abovecaptionskip}{0.cm}
        \includegraphics[width=0.9\linewidth]{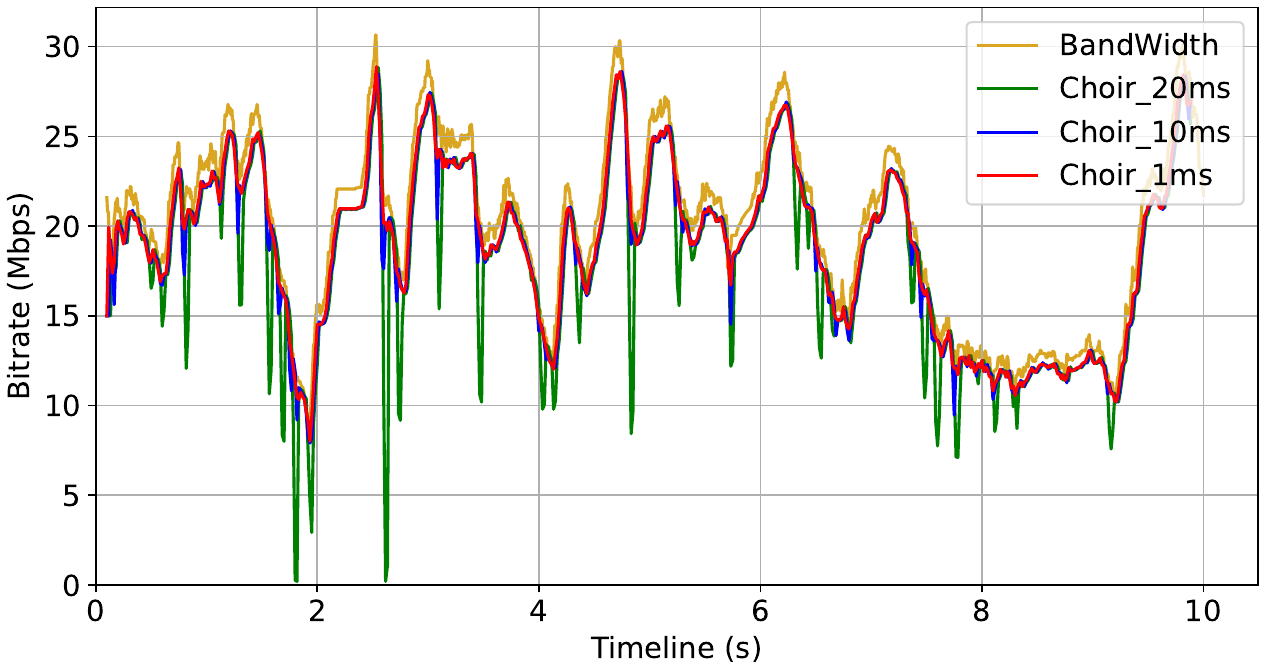}
        \caption{Frame bitrate variation.}
        \label{fig:bandwidth_choir_1_10_20ms}
    \end{minipage}  
    \vspace{-0.5cm} %
\end{figure}

\textbf{Impact of varied wired network latency.} Fig.~\ref{fig:bandwidth_choir_1_10_20ms} presents the bitrate variations of Choir under different wired network latency, alongside the bandwidth fluctuation trends from real 5G traces used in our simulation experiments. Choir could converge effectively to the available bandwidth at lower network latency, such as 1ms. 

\begin{figure}[! htbp]
    \centering
    \begin{minipage}{\linewidth}
        \centering
        \setlength{\abovecaptionskip}{0.cm}
        \includegraphics[width=\linewidth]{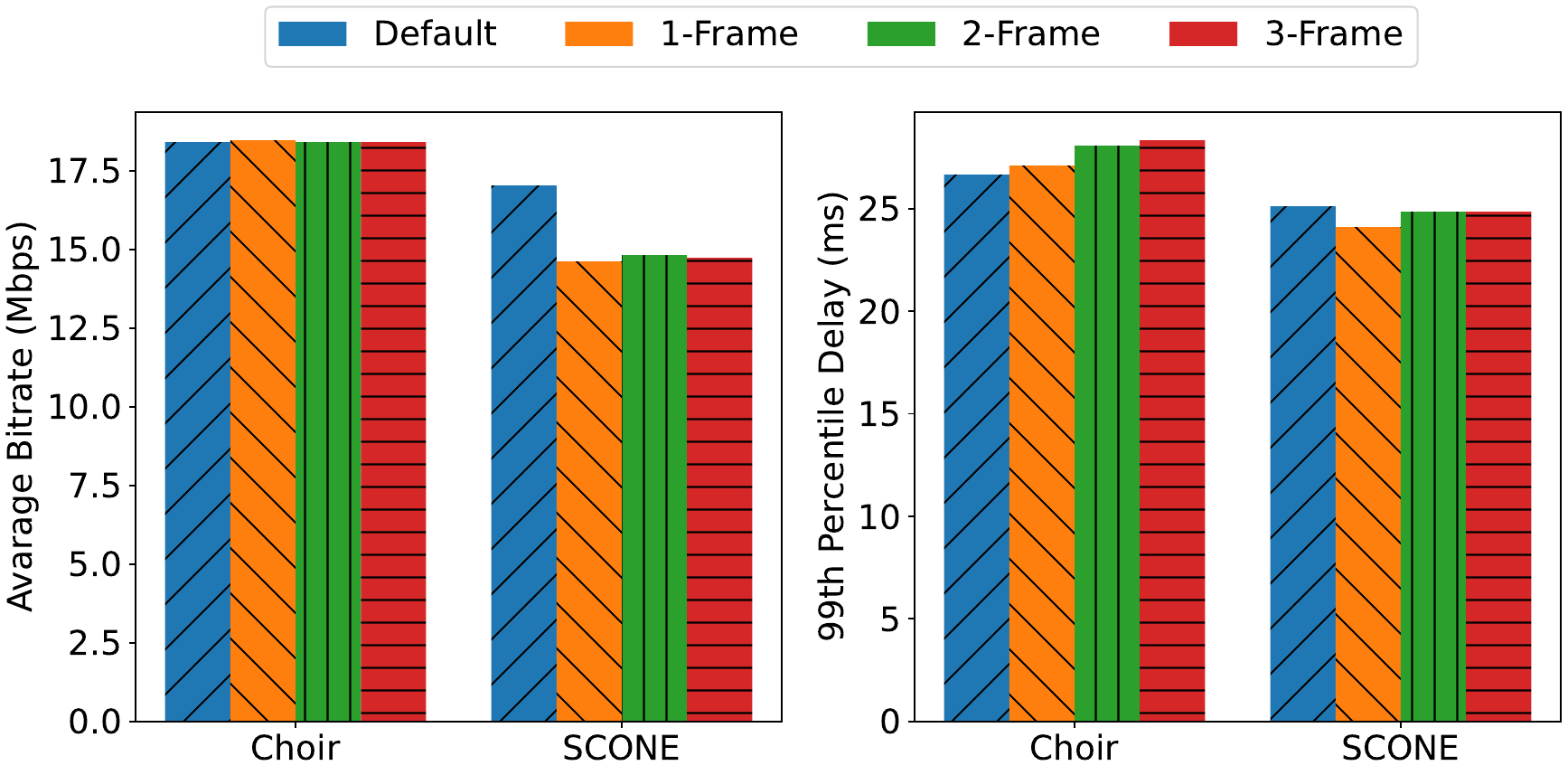}
        \caption{Comparison of different ACK frequencies }
        \label{fig:ack_f_duibi_choir_10ms}
    \end{minipage}   
    
\end{figure}

However, as wired network latency increases, Choir's bitrate fluctuations become more pronounced, reflecting the challenges introduced by increased feedback delay, which makes accurate predictions more difficult. 
Despite this, Choir strives to converge to the available bandwidth as closely as possible and keeps best performance among all compared solutions, demonstrating its adaptability to varying network conditions.

Considering that selective acknowledgement mechanisms may reduce the frequency and quantity of ACK packets, thereby degrading the feedback timeliness of base station , we evaluate Choir and SCONE under lower ACK intervals (one ACK per 1/2/3 frames).

As shown in the Fig.~\ref{fig:ack_f_duibi_choir_10ms}, Choir effectively adapts to different ACK frequencies, maintaining a relatively stable bitrate. 
Tail delay increases slightly in lower ACK frequency is also reasonable.

\begin{figure}[h] %
        \centering
        \begin{subfigure}[b]{0.49\linewidth}
            \centering
            \includegraphics[width=\textwidth]{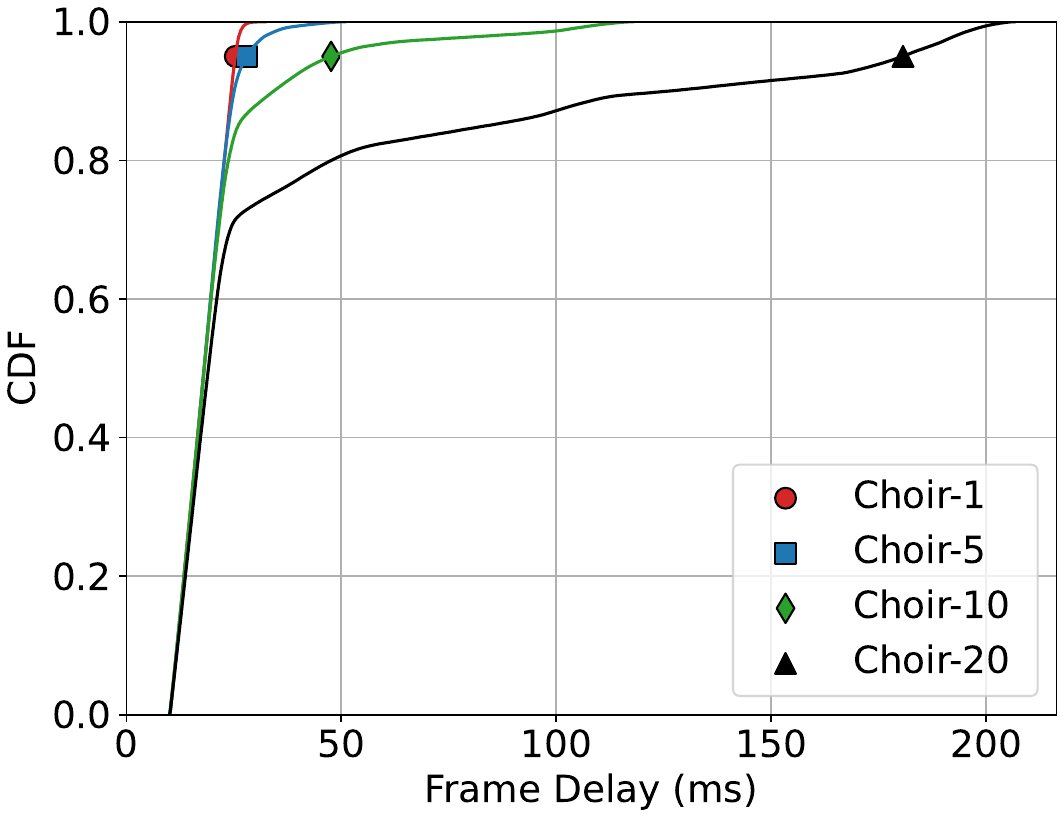} %
            \caption{Delay}
            \label{fig:pinghua_delay_cdf_choir_10ms}
        \end{subfigure}
        \begin{subfigure}[b]{0.49\linewidth}
            \centering
            \includegraphics[width=\textwidth]{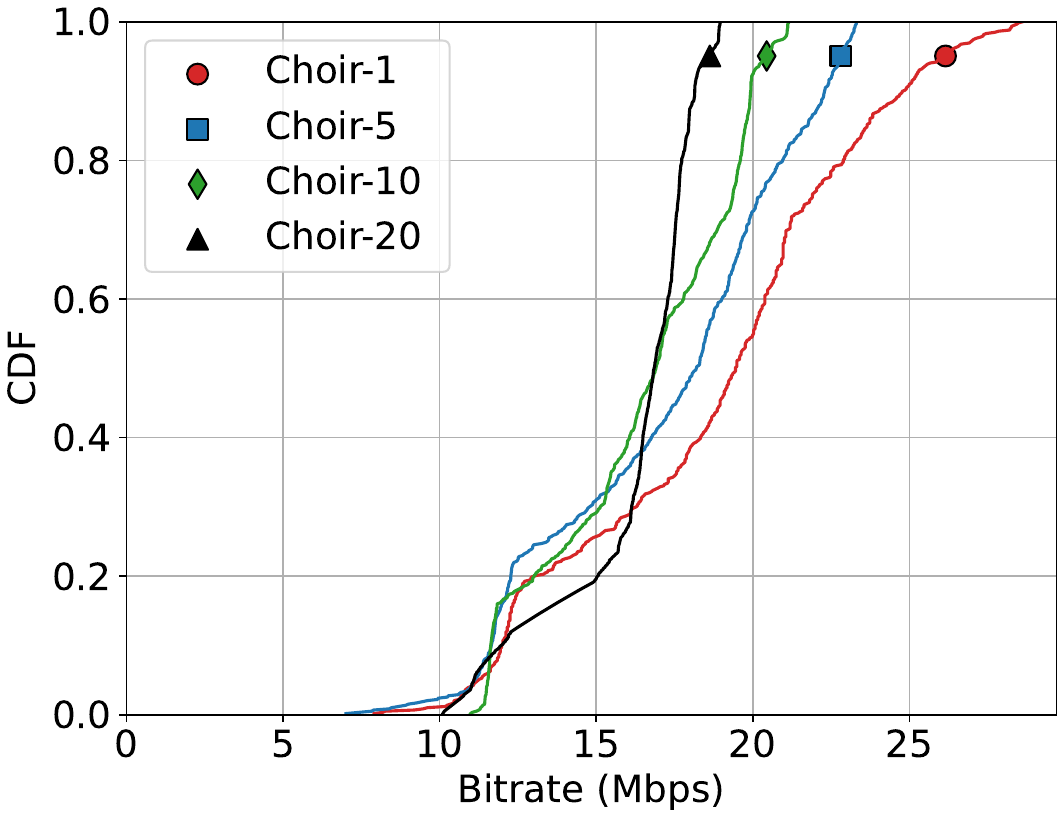} %
            \caption{Bitrate}
            \label{fig:pinghua_avgbitrate_caf_choir_10ms}
        \end{subfigure}
        \caption{The impact of smoothing factor.}
        \label{fig:pinghua}
\end{figure}

\textbf{Impact of smoothed bitrate control.}
We configure the smoothing factor of sender rate control algorithm to 1, 5, 10, and 20. As shown in Fig.~\ref{fig:pinghua}, as the smoothing factor increases, robustness improves, leading to smaller variations in video bitrate. 

However, this also results in a higher 99th percentile tail delay, indicating a trade-off between bitrate stability and delay performance.

\section{Discussion}
\textbf{Robust fairness under adversarial sender behavior.}
Base station inherently mitigates adversarial sender behaviors. 
Even if a sender bypasses Choir’s rate control algorithm, it cannot degrade other flows’ performance. 
The base station enforces balanced bandwidth allocation across flows. 
Senders exceeding allocated bandwidth by raising bitrates only incur higher tail delay, while reducing bitrates fails to meaningfully lower delay and may compromise throughput. 
Thus, we recommend senders strictly adhere to the guidance bandwidth to maximize capacity utilization and achieve optimal performance.

\textbf{Deployability.}
Both RTP streams and QUIC/TCP streams support IP options and do not require receiver-side participation for deployment. 
Many network-assisted solutions like ABC and L4S, necessitates involvement from the sender, network nodes, and receiver. Choir only requires recognition by the sender and base station, thereby enabling easier deployment.

\textbf{Overhead.} 
The computational overhead of Choir and L4S at the base station is approximately the same, as both conduct estimation based on bandwidth and queue conditions. 
However, Choir reduces communication overhead by returning directly to the server without requiring additional processing through the UE.
\section{Related Work}
Numerous explorations have been conducted regarding end-to-end low-latency transmission. 
Partial research focuses on sender-side optimization. 
For instance, Confucius \cite{meng2023confucius} enhances congestion control effectiveness by proactively moderating bandwidth allocation at access points, though this strategy exhibits weak correlation with RTBC. 
The AFR \cite{meng2023enabling} scheme further exploits the sender's capability to actively adapt to link state variations. 
From the client-side perspective, JitBright \cite{zhao2024jitbright} proposes improving real-time streaming experiences through jitter buffer optimization. 

For end-to-end latency minimization, diverse strategies have been explored. Approaches such as Copa, SQP, and Pudica effectively reduce latency through sophisticated end-to-end congestion detection techniques.
However, these methods remain constrained by reaction lag due to the inherent propagation delay between detection and response.

To address this, solutions like Hairpin \cite{meng2024hairpin} and Tambur \cite{rudow2023tambur} introduce Forward Error Correction (FEC) to mitigate packet loss, thereby indirectly improving latency performance, particularly in mobile network environments. Hairpin \cite{meng2024hairpin} effectively integrates FEC with link-layer retransmission mechanisms to mitigate data packet loss in wireless environments. 

Additionally, several works attempt to integrate deep reinforcement learning for video bitrate adaptation, including Learning Congestion Control \cite{zhou2019learning}, ONRL \cite{zhang2020onrl}, and LOKI \cite{zhang2021loki}. These methods inherit GCC’s limitations in fairness and robustness, and struggle to address cross-traffic fairness challenges.

\section{Conclusion}
Existing rate control solutions aiming to achieve the \textit{Performance Triangle} of RTBC often have to make tradeoff decisions among the three metrics: high throughput, low delay and fairness. 

The native dynamic delay caused by underlying architecture of 5G RAN is the root cause. But the native resource fairness allocation strategy provides us the opportunity to achieve the \textit{Performance Triangle} with the collaboration of 5G base station.

We proposes Choir, a innovative collaborative solution mainly deployed on 5G base stations that deeply integrates 5G radio characteristics and video streaming traffic patterns to guide sender-side efficient rate control.

Extensive simulation and testbed evaluations demonstrate Choir's significant performance on high average bitrate, low tail delay and inter-flow fairness over different 5G network scenarios.



\bibliographystyle{unsrt}
\bibliography{reference}

@inproceedings{vrwhitepaper,
  author = {HUAWEI-iLab},
  title = {CloudVR Solution White Paper},
  url = {https://www.huawei.com/en/news/2018/9/cloud-vr-solution-white-paper}
 
}

@techreport{ramakrishnan2001addition,
  title={The addition of explicit congestion notification (ECN) to IP},
  author={Ramakrishnan, Kadangode and Floyd, Sally and Black, David},
  year={2001}
}

@article{briscoe2019implementing,
  title={Implementing the’Prague Requirements’ for Low Latency Low Loss Scalable Throughput (L4S)},
  author={Briscoe, Bob and De Schepper, Koen and Tilmans, Olivier and K{\"u}hlewind, Mirja and Misund, Joakim and Albisser, Olga and Ahmed, A Sajjad},
  journal={Netdev 0x13},
  year={2019}
}

@misc{l4s,
  title={RFC 9330: Low Latency, Low Loss, and Scalable Throughput (L4S) Internet Service: Architecture},
  author={De Schepper, K and Bagnulo, M and White, G},
  year={2023},
  publisher={RFC Editor}
}

@inproceedings{zhuge,
  title={Achieving consistent low latency for wireless real-time communications with the shortest control loop},
  author={Meng, Zili and Guo, Yaning and Sun, Chen and Wang, Bo and Sherry, Justine and Liu, Hongqiang Harry and Xu, Mingwei},
  booktitle={Proceedings of the ACM SIGCOMM 2022 Conference},
  pages={193--206},
  year={2022}
}

@misc{johansson2017scream,
    series =    {Request for Comments},
    number =    8298,
    howpublished =  {RFC 8298},
    author =    {Ingemar Johansson and Zaheduzzaman Sarker},
    title =     {{Self-Clocked Rate Adaptation for Multimedia}},
    pagetotal = 36,
    year =      2017,
    month =     dec,
}

@article{ray2022sqp,
  title={SQP: Congestion Control for Low-Latency Interactive Video Streaming},
  author={Ray, Devdeep and Smith, Connor and Wei, Teng and Chu, David and Seshan, Srinivasan},
  journal={arXiv preprint arXiv:2207.11857},
  year={2022}
}

@article{son2024adaptable,
  title={Adaptable L4S Congestion Control for Cloud-Based Real-Time Streaming Over 5G},
  author={Son, Jangwoo and Sanchez, Yago and Hellge, Cornelius and Schierl, Thomas},
  journal={IEEE Open Journal of Signal Processing},
  year={2024},
  publisher={IEEE}
}

@inproceedings{carlucci2016analysis,
  title={Analysis and design of the google congestion control for web real-time communication (WebRTC)},
  author={Carlucci, Gaetano and De Cicco, Luca and Holmer, Stefan and Mascolo, Saverio},
  booktitle={Proceedings of the 7th International Conference on Multimedia Systems},
  pages={1--12},
  year={2016}
}

@inproceedings{arun2018copa,
  title={Copa: Practical $\{$Delay-Based$\}$ congestion control for the internet},
  author={Arun, Venkat and Balakrishnan, Hari},
  booktitle={15th USENIX Symposium on Networked Systems Design and Implementation (NSDI 18)},
  pages={329--342},
  year={2018}
}

@inproceedings{wang2024pudica,
  title={Pudica: Toward $\{$Near-Zero$\}$ Queuing Delay in Congestion Control for Cloud Gaming},
  author={Wang, Shibo and Yang, Shusen and Kong, Xiao and Wu, Chenglei and Jiang, Longwei and Xu, Chenren and Zhao, Cong and Yang, Xuesong and Xiao, Jianjun and Liu, Xin and others},
  booktitle={21st USENIX Symposium on Networked Systems Design and Implementation (NSDI 24)},
  pages={113--129},
  year={2024}
}

@inproceedings{goyal2020abc,
  title={$\{$ABC$\}$: A simple explicit congestion controller for wireless networks},
  author={Goyal, Prateesh and Agarwal, Anup and Netravali, Ravi and Alizadeh, Mohammad and Balakrishnan, Hari},
  booktitle={17th USENIX Symposium on Networked Systems Design and Implementation (NSDI 20)},
  pages={353--372},
  year={2020}
}

@techreport{li-6man-apn-ipv6-encap-00,
    number =    {draft-li-6man-apn-ipv6-encap-00},
    type =      {Internet-Draft},
    institution =   {Internet Engineering Task Force},
    publisher = {Internet Engineering Task Force},
    note =      {Work in Progress},
    url =       {https://datatracker.ietf.org/doc/draft-li-6man-apn-ipv6-encap/00/},
    author =    {Zhenbin Li and Shuping Peng and Chongfeng Xie and Shuai Zhang},
    title =     {{Application-aware IPv6 Networking (APN6) Encapsulation}},
    pagetotal = 14,
    year =      2024,
    month =     mar,
    day =       4,
}

@inproceedings{rudow2023tambur,
  title={Tambur: Efficient loss recovery for videoconferencing via streaming codes},
  author={Rudow, Michael and Yan, Francis Y and Kumar, Abhishek and Ananthanarayanan, Ganesh and Ellis, Martin and Rashmi, KV},
  booktitle={20th USENIX Symposium on Networked Systems Design and Implementation (NSDI 23)},
  pages={953--971},
  year={2023}
}

@inproceedings{meng2024hairpin,
  title={Hairpin: Rethinking packet loss recovery in edge-based interactive video streaming},
  author={Meng, Zili and Kong, Xiao and Chen, Jing and Wang, Bo and Xu, Mingwei and Han, Rui and Liu, Honghao and Arun, Venkat and Hu, Hongxin and Wei, Xue},
  booktitle={21st USENIX Symposium on Networked Systems Design and Implementation (NSDI 24)},
  pages={907--926},
  year={2024}
}

@misc{xgproduct,
  title = {{{xgproduct.}}},
  author = {WITCOMM.},
  year = {2023},
  howpublished = {\url{https://witcomm.net/xgstation}}
}

@techreport{3gpp.23.288,
  author = {{3GPP}},
  title = {{3GPP TS 23.288}: Architecture enhancements for 5G System (5GS) to support network data analytics services},
  number = {23.288},
  type = {Technical Specification (TS)},
  institution = {3rd Generation Partnership Project (3GPP)},
  year = {2019},
  month = {December},
  note = {Version 15.4.0},
  url = {https://www.3gpp.org/ftp/Specs/archive/23_series/23.288/23288-f40.zip},
  urldate = {2022-03-17}
}

@misc{3gpp-release-16,
  title = {{{Release 16}}},
  author = {3GPP},
  year = {2024},
  howpublished = {\url{https://www.3gpp.org/specifications-technologies/releases/release-16}}
}

@article{torres2020immersive,
  title={Immersive interconnected virtual and augmented reality: A 5G and IoT perspective},
  author={Torres Vega, Maria and Liaskos, Christos and Abadal, Sergi and Papapetrou, Evangelos and Jain, Akshay and Mouhouche, Belkacem and Kalem, G{\"o}khan and Erg{\"u}t, Salih and Mach, Marian and Sabol, Tomas and others},
  journal={Journal of Network and Systems Management},
  volume={28},
  pages={796--826},
  year={2020},
  publisher={Springer}
}

@article{rubio2017immersive,
  title={Immersive environments and virtual reality: Systematic review and advances in communication, interaction and simulation},
  author={Rubio-Tamayo, Jose Luis and Gertrudix Barrio, Manuel and Garc{\'\i}a Garc{\'\i}a, Francisco},
  journal={Multimodal technologies and interaction},
  volume={1},
  number={4},
  pages={21},
  year={2017},
  publisher={MDPI}
}

@inproceedings{kamarainen2017measurement,
  title={A measurement study on achieving imperceptible latency in mobile cloud gaming},
  author={K{\"a}m{\"a}r{\"a}inen, Teemu and Siekkinen, Matti and Yl{\"a}-J{\"a}{\"a}ski, Antti and Zhang, Wenxiao and Hui, Pan},
  booktitle={Proceedings of the 8th ACM on Multimedia Systems Conference},
  pages={88--99},
  year={2017}
}

@ARTICLE{7938641,
  author={Carlucci, Gaetano and De Cicco, Luca and Holmer, Stefan and Mascolo, Saverio},
  journal={IEEE/ACM Transactions on Networking}, 
  title={Congestion Control for Web Real-Time Communication}, 
  year={2017},
  volume={25},
  number={5},
  pages={2629-2642},
  doi={10.1109/TNET.2017.2703615}}

@inproceedings{singh2013performance,
  title={Performance analysis of receive-side real-time congestion control for WebRTC},
  author={Singh, Varun and Lozano, Albert Abello and Ott, Jorg},
  booktitle={2013 20th International Packet Video Workshop},
  pages={1--8},
  year={2013},
  organization={IEEE}
}

@article{kaltenberger2020openairinterface,
  title={OpenAirInterface: Democratizing innovation in the 5G Era},
  author={Kaltenberger, Florian and Silva, Aloizio P and Gosain, Abhimanyu and Wang, Luhan and Nguyen, Tien-Thinh},
  journal={Computer Networks},
  volume={176},
  pages={107284},
  year={2020},
  publisher={Elsevier}
}

@inproceedings{xie2020pbe,
  title={PBE-CC: Congestion control via endpoint-centric, physical-layer bandwidth measurements},
  author={Xie, Yaxiong and Yi, Fan and Jamieson, Kyle},
  booktitle={Proceedings of the Annual conference of the ACM Special Interest Group on Data Communication on the applications, technologies, architectures, and protocols for computer communication},
  pages={451--464},
  year={2020}
}

@article{zhu2020NADA,
  title={Network-assisted dynamic adaptation (NADA): a unified congestion control scheme for real-time media},
  author={Zhu, Xiaoqing and Pan, Rong and Ramalho, M and Mena, S},
  journal={RFC 8698},
  year={2020}
}

@inproceedings{zhang2019congestion,
  title={Congestion control for RTP media: A comparison on simulated environment},
  author={Zhang, Songyang and Lei, Weimin and Zhang, Wei and Guan, Yunchong},
  booktitle={International Conference on Simulation Tools and Techniques},
  pages={43--52},
  year={2019},
  organization={Springer}
}

@article{patriciello2019e2e,
  title={An E2E simulator for 5G NR networks},
  author={Patriciello, Natale and Lagen, Sandra and Bojovic, Biljana and Giupponi, Lorenza},
  journal={Simulation Modelling Practice and Theory},
  volume={96},
  pages={101933},
  year={2019},
  publisher={Elsevier}
}

@article{bbr,
  title={BBR: Congestion-based congestion control: Measuring bottleneck bandwidth and round-trip propagation time},
  author={Cardwell, Neal and Cheng, Yuchung and Gunn, C Stephen and Yeganeh, Soheil Hassas and Jacobson, Van},
  journal={ACM Queue},
  volume={14},
  number={5},
  pages={20--53},
  year={2016},
  publisher={ACM New York, NY, USA}
}

@article{cubic,
  title={CUBIC: a new TCP-friendly high-speed TCP variant},
  author={Ha, Sangtae and Rhee, Injong and Xu, Lisong},
  journal={ACM SIGOPS operating systems review},
  volume={42},
  number={5},
  pages={64--74},
  year={2008},
  publisher={ACM New York, NY, USA}
}

@inproceedings{dddsu,
  title={Performance evaluation of extended reality applications in 5g nr system},
  author={Sundararajan, Jay Kumar and Kwon, Hwan-Joon and Awoniyi-Oteri, Olufunmilola and Kim, Yuchul and Li, Chih-Ping and Damnjanovic, Jelena and Zhou, Shanyu and Ma, Ruifeng and Tokgoz, Yeliz and Hande, Prashanth and others},
  booktitle={2021 PIMRC},
  pages={1--7},
  year={2021},
  organization={IEEE}
}

@article{lena,
  author    = {Katerina Koutlia and
               Biljana Bojovic and
               Zoraze Ali and
               Sandra Lag{\'{e}}n},
  title     = {Calibration of the 5G-LENA system level simulator in 3GPP reference
               scenarios},
  journal   = {Simul. Model. Pract. Theory},
  volume    = {119},
  pages     = {102580},
  year      = {2022},
}

@inproceedings{meng2023enabling,
  title={Enabling High Quality $\{$Real-Time$\}$ Communications with Adaptive $\{$Frame-Rate$\}$},
  author={Meng, Zili and Wang, Tingfeng and Shen, Yixin and Wang, Bo and Xu, Mingwei and Han, Rui and Liu, Honghao and Arun, Venkat and Hu, Hongxin and Wei, Xue},
  booktitle={20th USENIX Symposium on Networked Systems Design and Implementation (NSDI 23)},
  pages={1429--1450},
  year={2023}
}

@inproceedings{lecci2021ns,
  title={An ns-3 implementation of a bursty traffic framework for virtual reality sources},
  author={Lecci, Mattia and Zanella, Andrea and Zorzi, Michele},
  booktitle={Proceedings of the 2021 Workshop on ns-3},
  pages={73--80},
  year={2021}
}

@inproceedings{xquic3,
title = {{{XQUIC Library released by Alibaba is a cross-platform implementation of QUIC and HTTP/3 protocol.}}},
  author = {Alibaba.},
  year = {2022},
  howpublished = {\url{https://github.com/alibaba/xquic}}
}

@inproceedings{zhou2019learning,
  title={Learning to coordinate video codec with transport protocol for mobile video telephony},
  author={Zhou, Anfu and Zhang, Huanhuan and Su, Guangyuan and Wu, Leilei and Ma, Ruoxuan and Meng, Zhen and Zhang, Xinyu and Xie, Xiufeng and Ma, Huadong and Chen, Xiaojiang},
  booktitle={The 25th Annual International Conference on Mobile Computing and Networking},
  pages={1--16},
  year={2019}
}

@misc{TS38331,
  author       = {3rd Generation Partnership Project (3GPP)},
  title        = {Technical Specification Group Radio Access Network; NR; Radio Resource Control (RRC) protocol specification},
  howpublished = {3GPP TS 38.331},
  year         = {2023},
  month        = sep,
  note         = {Version 17.7.0},
  url          = {https://www.3gpp.org/ftp/Specs/archive/38_series/38.331/38331-g70.zip}
}

@misc{TS38322,
  author       = {3rd Generation Partnership Project (3GPP)},
  title        = {Technical Specification Group Radio Access Network; NR; Radio Link Control (RLC) protocol specification},
  howpublished = {3GPP TS 38.322},
  year         = {2023},
  month        = sep,
  note         = {Version 17.7.0},
  url          = {https://www.3gpp.org/ftp/Specs/archive/38_series/38.322/38322-g70.zip}
}

@misc{openairinterface5g,
  author       = {OpenAirInterface},
  title        = {OpenAirInterface 5G: Feature Set Documentation},
  howpublished = {\url{https://gitlab.eurecom.fr/oai/openairinterface5g/blob/develop/doc/FEATURE_SET.md}},
  note         = {Accessed: 2024-10-09}
}

@techreport{ietf-avtcore-rtp-over-quic-11,
    number =    {draft-ietf-avtcore-rtp-over-quic-11},
    type =      {Internet-Draft},
    institution =   {Internet Engineering Task Force},
    publisher = {Internet Engineering Task Force},
    note =      {Work in Progress},
    url =       {https://datatracker.ietf.org/doc/draft-ietf-avtcore-rtp-over-quic/11/},
    author =    {Mathis Engelbart and Joerg Ott and Spencer Dawkins},
    title =     {{RTP over QUIC (RoQ)}},
    pagetotal = 75,
    year =      2024,
    month =     jul,
    day =       8,
}

@inproceedings{brown2024principles,
  title={Principles for internet congestion management},
  author={Brown, Lloyd and Alcoz, Albert Gran and Cangialosi, Frank and Narayan, Akshay and Alizadeh, Mohammad and Balakrishnan, Hari and Friedman, Eric and Katz-Bassett, Ethan and Krishnamurthy, Arvind and Schapira, Michael and others},
  booktitle={Proceedings of the ACM SIGCOMM 2024 Conference},
  pages={166--180},
  year={2024}
}

@techreport{shi-scone-rtc-requirement-02,
    number =    {draft-shi-scone-rtc-requirement-02},
    type =      {Internet-Draft},
    institution =   {Internet Engineering Task Force},
    publisher = {Internet Engineering Task Force},
    note =      {Work in Progress},
    url =       {https://datatracker.ietf.org/doc/draft-shi-scone-rtc-requirement/02/},
    author =    {Hang Shi and Xuesong Geng and Qiangzhou Gao and Qinghua Wu and Jiaxing Zhang},
    title =     {{SCONE Real Time Communication Requirement}},
    pagetotal = 6,
    year =      2025,
    month =     apr,
    day =       20,
}

@article{sentosacellreplay,
  title={CellReplay: Towards accurate record-and-replay for cellular networks},
  author={Sentosa, William and Chandrasekaran, Balakrishnan and Godfrey, P Brighten and Hassanieh, Haitham},
  booktitle={22nd USENIX Symposium on Networked Systems Design and Implementation (NSDI 25)},
  year={NSDI 2025},
}

@inproceedings{zhang2020onrl,
  title={OnRL: Improving mobile video telephony via online reinforcement learning},
  author={Zhang, Huanhuan and Zhou, Anfu and Lu, Jiamin and Ma, Ruoxuan and Hu, Yuhan and Li, Cong and Zhang, Xinyu and Ma, Huadong and Chen, Xiaojiang},
  booktitle={Proceedings of the 26th Annual International Conference on Mobile Computing and Networking},
  pages={1--14},
  year={2020}
}

@inproceedings{zhang2021loki,
  title={Loki: improving long tail performance of learning-based real-time video adaptation by fusing rule-based models},
  author={Zhang, Huanhuan and Zhou, Anfu and Hu, Yuhan and Li, Chaoyue and Wang, Guangping and Zhang, Xinyu and Ma, Huadong and Wu, Leilei and Chen, Aiyun and Wu, Changhui},
  booktitle={Proceedings of the 27th Annual International Conference on Mobile Computing and Networking},
  pages={775--788},
  year={2021}
}

@inproceedings{zhao2024jitbright,
  title={JitBright: towards Low-Latency Mobile Cloud Rendering through Jitter Buffer Optimization},
  author={Zhao, Yuankang and Wu, Qinghua and Lv, Gerui and Yang, Furong and Zhang, Jiuhai and Peng, Feng and Liu, Yanmei and Li, Zhenyu and Chen, Ying and Guo, Hongyu and others},
  booktitle={Proceedings of the 34th edition of the Workshop on Network and Operating System Support for Digital Audio and Video},
  pages={36--42},
  year={2024}
}

@article{meng2023confucius,
  title={Confucius: Achieving Consistent Low Latency with Practical Queue Management for Real-Time Communications},
  author={Meng, Zili and Atre, Nirav and Xu, Mingwei and Sherry, Justine and Apostolaki, Maria},
  journal={arXiv preprint arXiv:2310.18030},
  year={2023}
}

\appendix 

\section{Openxg bandwidth trace and b210 bandwidth trace} 
\begin{figure}[htbp
]
    \centering
        \centering
        \setlength{\abovecaptionskip}{0.cm}
        \includegraphics[width=0.9\linewidth]{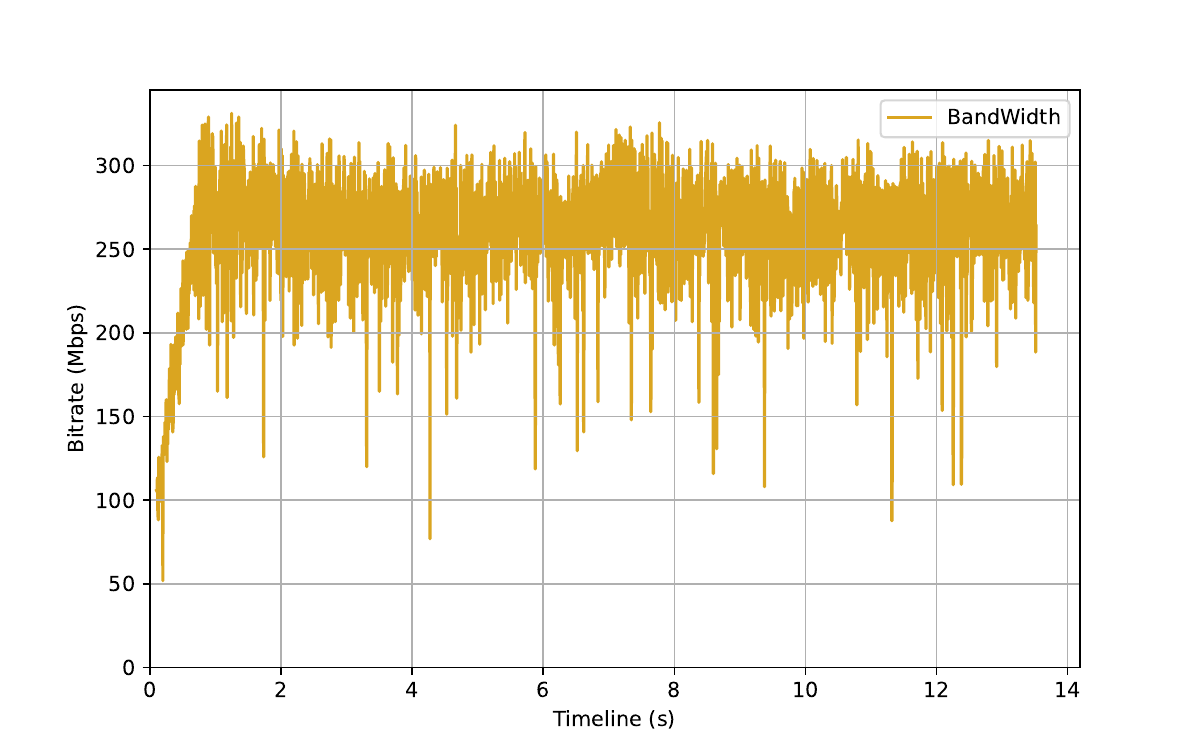}
        \caption{Openxg bandwidth trace.}
        \label{}
\end{figure}

\begin{figure}[htbp]
    \centering
        \centering
        \setlength{\abovecaptionskip}{0.cm}
        \includegraphics[width=0.9\linewidth]{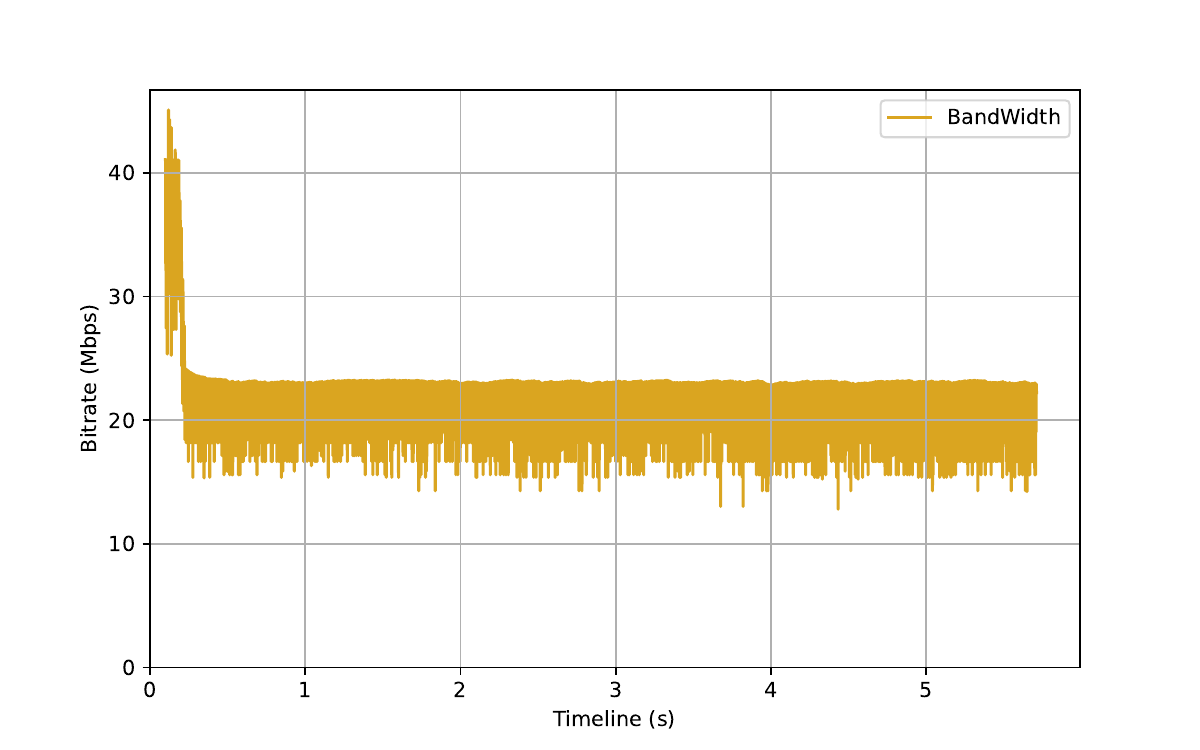}
        \caption{B210 bandwidth trace.}
        \label{fig:fdcc_1_10_20ms_duibi}
\end{figure}

\end{document}